\documentclass[10pt,journal]{IEEEtran}
\usepackage{ifpdf}
\usepackage{amssymb}
\usepackage{color}

\usepackage{multirow}
\usepackage{cite}

\ifCLASSINFOpdf
  \usepackage[pdftex]{graphicx}
  \graphicspath{{../pdf/}{../jpeg/}}
  \DeclareGraphicsExtensions{.pdf,.jpeg,.png}
\else
\usepackage[dvips]{graphicx}
\usepackage{diagbox}
\usepackage{slashbox}
\fi
\usepackage{amsmath}
\usepackage{algorithmic}

\usepackage{array}
\usepackage[justification=centering,font=footnotesize]{caption}

\ifCLASSOPTIONcompsoc
  \usepackage[caption=false,font=normalsize,labelfont=sf,textfont=sf]{subfig}
\else
  \usepackage[caption=false,font=footnotesize]{subfig}
\fi

\usepackage{booktabs}
\usepackage{fixltx2e}
\usepackage{multicol}

\usepackage{stfloats}
\usepackage{url}

\begin{document}
%
\title{$SMAPGAN$: Generative Adversarial Network-Based Semi-Supervised Styled Map Tile Generation Method}
%
%
%

\author{Xu~Chen, Songqiang~Chen, Tian~Xu, Bangguo~Yin, Jian~Peng, Xiaoming~Mei, and~Haifeng~Li~\IEEEmembership{Member,~IEEE}
\thanks{}
\thanks{\emph{Corresponding author:~Haifeng~Li~(email:lihaifeng@csu.edu.cn)}}
\thanks{Xu~Chen, Songqiang~Chen, Tian~Xu,and~Bangguo~Yin are with School of Computer Science, Wuhan University, Wuhan, 430072, China.}
\thanks{Jian~Peng, Xiaoming~Mei and Haifeng Li are with School of Geosciences and Info-Physics, Central South University, Changsha, 410083, China.}
}

%
%

\markboth{IEEE TRANSACTIONS ON GEOSCIENCE AND REMOTE SENSING}
{Shell \MakeLowercase{\textit{et al.}}: $SMAPGAN$: Generative Adversarial Network-Based Semi-Supervised Styled Map Tile Generation Method}
%



\maketitle

\begin{abstract}
Traditional online map tiles, which are widely used on the Internet, such as by Google Maps and Baidu Maps, are rendered from vector data. The timely updating of online map tiles from vector data, for which generation is time consuming, is a difficult mission. Generating map tiles over time from remote sensing images is relatively simple and can be performed quickly without vector data. However, this approach used to be challenging or even impossible. Inspired by image-to-image translation (img2img) techniques based on generative adversarial networks (GANs), we proposed a semi-supervised generation of styled map tiles based on GANs (SMAPGAN) model to generate styled map tiles directly from remote sensing images. In this model, we designed a semi-supervised learning strategy to pre-train SMAPGAN on rich unpaired samples and fine-tune it on limited paired samples in reality. We also designed the image gradient L1 loss and the image gradient structure loss to generate a styled map tile with global topological relationships and detailed edge curves for objects, which are important in cartography. Moreover, we proposed the edge structural similarity index (ESSI) as a metric to evaluate the quality of the topological consistency between the generated map tiles and ground truth. The experimental results show that SMAPGAN outperforms state-of-the-art (SOTA) works according to the mean squared error, the structural similarity index, and the ESSI. Also, SMAPGAN gained higher approval than SOTA in a human perceptual test on the visual realism of cartography. Our work shows that SMAPGAN is a new tool with excellent potential for producing styled map tiles. Our implementation of SMAPGAN is available at \url{https://github.com/imcsq/SMAPGAN}.
\end{abstract}
\begin{IEEEkeywords}
generative adversarial networks, map tiles generation, semi-supervised, quality assessment.
\end{IEEEkeywords}
\section{Introduction}
\IEEEPARstart{O}{nline} map services are widely used in several fields and the demand for high-quality online maps is increasing dramatically, as they are an important part of location services. Traditional manual cartography relies on vectorization of the map and is time consuming\cite{haunold_keystroke_1993}, which makes the timely update of object features on maps difficult. However, remote sensing images can be acquired in a timely manner. Thus, generating styled map tiles directly from remote sensing images without vector data is one method of reducing the time cost.

Although it used to be thought of as a challenging or even an impossible mission\cite{kang_transferring_2019}, the recent rise of image-to-image translation (img2img) techniques based on generative adversarial networks (GANs) offers a method for generating styled map tiles directly from remote sensing images because it can be viewed as an img2img task.

In recent years, GAN and img2img techniques have been used to produce images for various purposes, such as super-resolution\cite{han_modified_2019,jiang_edge-enhanced_2019,haut_new_2018}, image classification\cite{wang_enhancing_2018,zhu_generative_2018,wang_caps-triplegan:_2019}, dataset augmentation\cite{uricar_yes_2019,frid-adar_synthetic_2018,zhang_unsupervised_2019}, and generation of images in a specific domain\cite{wu_gp-gan:_2019,tran_extreme_2018,zhang_deeproad:_2018,suarez_deep_2018}.

The img2img technique and its application have developed quickly. First, Pix2Pix\cite{isola_image--image_2017} and CycleGAN\cite{zhu_unpaired_2017} were put forward as two typical img2img techniques. Then, \cite{liu_unsupervised_2017} and \cite{alami_mejjati_unsupervised_2018} tried to improve generating quality by adding a shared latent space and an attention layer to CycleGAN\cite{zhu_unpaired_2017}, respectively. These img2img models were applied to generate multi-styled maps from simple-styled maps, which were derived from vector data \cite{kang_transferring_2019} and map tiles from remote sensing images \cite{ganguli_geogan:_2019}. However, these generated maps suffered erroneous topological relations and blurred details. The current flaws of applying general img2img in map generation from remote sensing images lay in three aspects.

First, the lack of paired samples consisting of a remote sensing image and its corresponding map tile makes the results inadequate. For example, the latest remote sensing images may lack corresponding online map tiles, while online map tiles may be generated from other methods without the corresponding remote sensing images. However, unpaired data could increase the diversity of object features in order to improve the generalization ability of the model.

Second, although several studies have the pixel-wise L1 distance between the output and the ground truth to improve the models \cite{wu_gp-gan:_2019,isola_image--image_2017,zhu_unpaired_2017,demir_patch-based_2018}, these studies ignored the topological relationship among objects, which is vital in cartography \cite{neumann_topological_1994}.

Third, the metric for evaluating the quality of generated map tiles is lacking. A few metrics have been proposed to evaluate the realism of the images generated by GAN \cite{salimans_improved_2016}, \cite{heusel_gans_2017}, but a high-quality map should not only be of a high realism, that it should also be able to replicate the correct features on the ground truth. Instead, full-reference metrics provide a pixel-wise comparison between the generated image and the ground truth, but no metrics focus on the topological relationship of objects.

In this paper, we designed a semi-supervised learning strategy to fit the model to both paired and unpaired samples. We compared SMAPGAN with SOTA baselines under the mean square error (MSE), the structural similarity index (SSIM), the ESSI, and a human perception test. The results show that SMAPGAN outperforms SOTA when performing styled map tiles generation tasks.

Our main ideas and contributions are listed as follows:

(1) We proposed a novel styled map tiles generation model named SMAPGAN, which introduced a semi-supervised learning strategy to train the model on massive unpaired samples and fewer paired samples.

(2) We designed an image gradient L1 loss and an image gradient structure loss to transform the topological relationship among the objects on maps to the topological relationship among the edge curves of the objects. These encouraged SMAPGAN to generate maps with a better topological relationship and better details than the baselines.

(3) We proposed a full-reference image quality metric, the edge structural similarity index (ESSI), to measure the topological consistency between the generated and the real maps using the correlation between random variables, to which the edge images of the generated and real maps are transformed.

This paper is structured as follows. In Section 2, the related works are presented. In Section 3, the SMAPGAN framework and its losses are discussed in detail. Specifically, the two proposed gradient losses are introduced. The experiment and results are discussed in Section 4. Finally, the advantages and shortcomings of our model as well as our future work are mentioned in Section 5.

\section{Related Work}
We divided the current related work into four aspects. Previous studies have proposed several basic img2img translation methods. Then other studies modified the basic model to obtain good images, while still being for general usage. Recently, studies have tried to use img2img methods to generate maps. Additionally, evaluating GAN is a prominent topic that inspires more research.

\subsection{Image to Image Translation}Pix2Pix \cite{isola_image--image_2017} is a typical case of an image to image translation that adopts a conditional GAN (cGAN) \cite{mirza_conditional_2014} to learn the translation in a supervised way. In the Pix2Pix model, a generator network is trained to produce valid and realistic images in the target domain corresponding to the input images from the source domain. This process is constrained by an adversarial and an L1 loss, which are considered to be sufficient high- and low-frequency structure restrictions, respectively. However, Pix2Pix must be trained on paired samples due to its supervised loss. To break these constraints, \cite{zhu_unpaired_2017} proposed CycleGAN by imposing cycle-consistency, which was built on the fact that an image should be reconstructed correctly after translating twice. This premise allows img2img models to be trained using unpaired datasets. However, the lack of ground truths makes the method less robust than a supervised one. These classical img2img translating models are the bases of our method.

\subsection{Augmented Image to Image Translating}After the two classic types of img2img models were proposed, recent works attempted to improve the results of the models for either a universal or specific purpose. \cite{liu_unsupervised_2017,huang_multimodal_2018,zhu_toward_2017,zhang_stackgan:_2017,choi_stargan:_2018,almahairi_augmented_2018} focused on the impact of latent features and modified the architecture of the proposed model to make improved, multiple, or other amazing results. Others \cite{alami_mejjati_unsupervised_2018,you_bayesian_2018,januszewski_segmentation-enhanced_2019,engin_cycle-dehaze:_2018,lu_guiding_2019} have advocated supplementing helpful components to improve the image generation, such as adding an attention layer into the model. In our work, we combined an important component with the structure of proposed models to improve the performance on styled map tiles generation.

\subsection{Latest Attempt of Using img2img technique to Generate Styled Map Tiles}With the popularity of img2img, studies have tried applying this method to generate styled map tiles. \cite{kang_transferring_2019} used Pix2Pix and CycleGAN to transfer styled map tiles from simple-styled maps generated based on vectorized data. \cite{ganguli_geogan:_2019} put forward a model based on cGAN with style and content losses to generate styled map tiles from satellite images. Both works verified the basic feasibility of the task and left a few quality-related problems, such as incorrect topological relations of the objects.

\subsection{Metrics of Img2Img Translation Work}Metrics, such as the inception score (IS) \cite{salimans_improved_2016} and Frechet inception distance (FID) \cite{heusel_gans_2017}, were proposed to evaluate the quality of images generated by GAN. However, the IS and FID are not suitable for map translation tasks. These metrics are generally used to measure an image-generative model based on the distance of the probability distribution between ground truths and generated images. The distance of probability cannot measure the quality of one single generated map tile compared to its ground truth because the map tiles translation demands a generated map tile to replicate the details of the ground truth. In geoscience, \cite{baraldi_quality_2005} proposed a quality assessment for classification and cluster maps, which compared the labeling and segmentation among cluster maps. However, this method cannot measure the accuracy of generated styled map tiles. 

Instead, full-reference image quality assessment indexes, such as MSE \cite{sammut_encyclopedia_2011} and SSIM \cite{wang_image_2004}, are suitable for assessing the map translation tasks. MSE measures the global average L2 distance of pixels between two images, but cannot measure the pixel-wise structure. In contrast to MSE, SSIM considers a pixel-wise structure, which can evaluate structural replication of generated map tiles. \cite{xue_gradient_2014} and \cite{ding_image_2017} further investigated the significance of the first-order derivative of pixels of images as well as the gradient of images to image quality assessment. This method inspired our work on assessing the quality of generated map tiles.

\section{Methods}
\subsection{SMAPGAN Framework}
\begin{figure*}[htbp]
\centering
\includegraphics[scale=0.6]{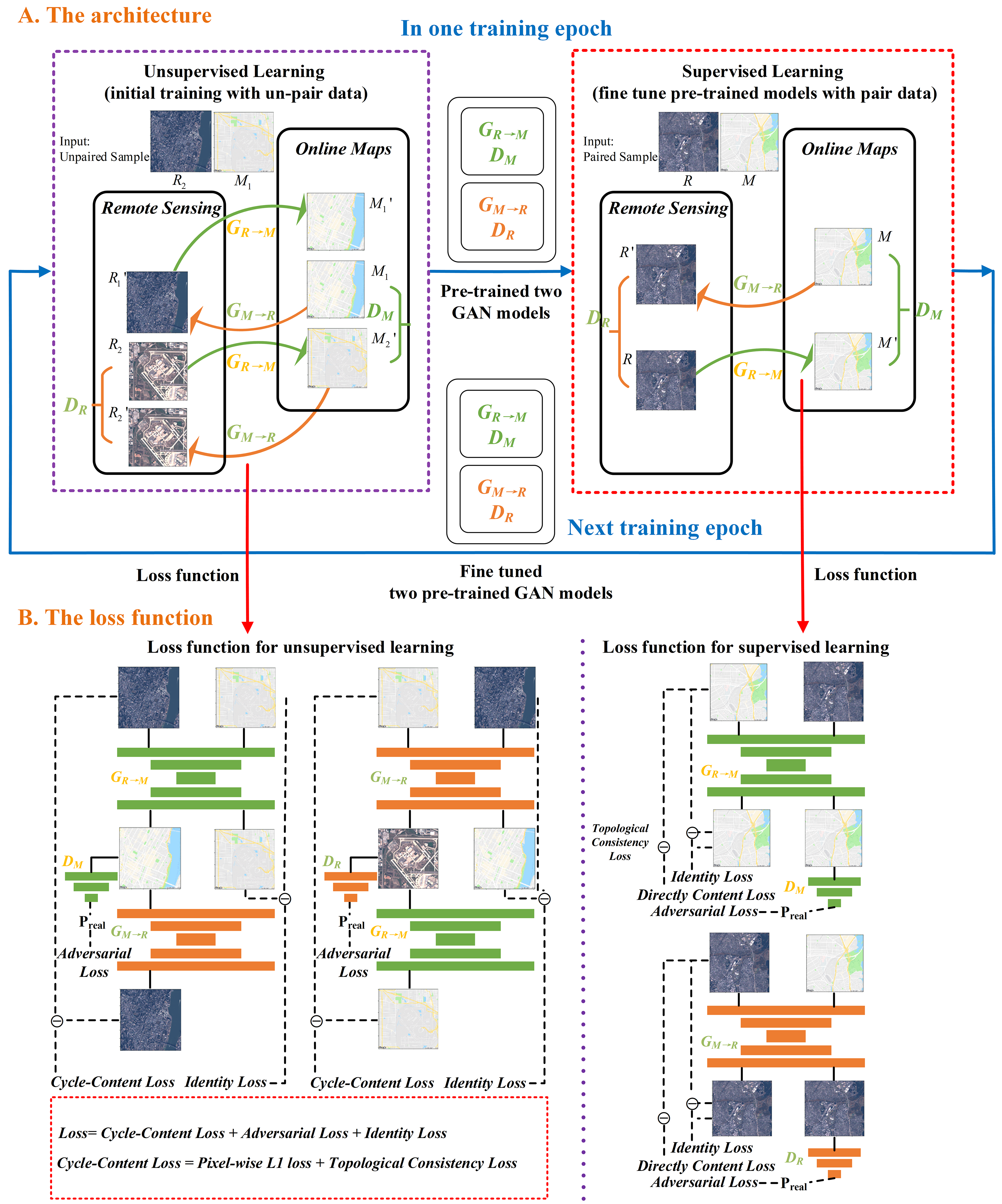}
\caption{A SMAPGAN Method Overview; B Unsupervised learning process and supervised learning process in detail.}
\label{Figure 1}
\end{figure*}
The SMAPGAN model includes two GANs that generate styled map tiles from input remote sensing images ($G_{R\rightarrow M},\ D_M$) (where $G_{R\rightarrow M}$ is the generator producing map tiles from remote sensing images, and $D_M$ is the discriminator that discriminates against generated map tiles) and generate remote sensing images from styled map tiles $(G_{M\rightarrow R},D_R)$ (where $G_{M\rightarrow R}$ is the generator producing remote sensing images from styled map tiles, and $D_R$ is the discriminator to discriminate against generated remote sensing images). As shown in Figure \ref{Figure 1}, SMAPGAN achieves semi-supervised learning by combining a supervised and an unsupervised learning strategy, which will be discussed in Section 3.2. Moreover, SMAPGAN focuses not only on the correct value of each pixel on the generated maps, but also on the quality of the topological relationship on the output, which will be discussed in detail in Section 3.3.

The architecture of SMAPGAN is shown in Figure \ref{Figure 1}(A). For each epoch in training, there are two stages:

1. Unsupervised pre-training stage. In this stage, we first train the model on unpaired samples with unsupervised learning to initially train the weight of the generators and discriminators. A remote sensing image or a map tile from unpaired samples is sent into a generator ($G_{R\rightarrow M}$ or $G_{M\rightarrow R}$) and the generator outputs a generated image (a map tile or a remote sensing image). Then the output image is sent in to the other generator ($G_{M\rightarrow R}$ or $G_{R\rightarrow M}$) to generate an image in the same domain as the raw input. This output after two translation is then sent into the discriminator ($D_R$ or $D_M$) to be discriminated. The weight of the generators and discriminators is trained by this process.

2. Supervised fine-tuning stage. After the pre-training stage on all unpaired samples, paired samples are sent into generators ($G_{R\rightarrow M}$ or $G_{M\rightarrow R}$) according to the type of the inputted image. The generator outputs an image in the target domain and the output is sent into the discriminator ($D_M$ or $D_R$) to be discriminated. The weight of the generators and discriminators is then fine-tuned by this process.

After these two stages, an epoch ends and another epoch begins with the same stages. Thus, we trained the model in a semi-supervised way.

\subsection{Loss Function in Semi-Supervised Learning}
Given the existence of various types of objects on maps and the fact that there are not always many paired samples available, we needed to train the model with available paired and unpaired samples, as described in Section 1. Therefore, we designed a semi-supervised learning strategy that combined unsupervised and supervised learning, as shown in Figure \ref{Figure 1}(A). This allowed us to train our model on paired and unpaired training samples.

As an img2img translation work, we constrained the model learning with three classic types of loss, namely, content loss, adversarial loss, and identity loss, \eqref{eq1} as did previous works \cite{isola_image--image_2017,zhu_unpaired_2017}.

\begin{align}
\label{eq1}
\mathcal{L}=\lambda_{ctn}\ \mathcal{L}_{ctn}+\lambda_{adv}\ \mathcal{L}_{adv}+\lambda_{idt}\ \mathcal{L}_{idt}   
\end{align}
\subsubsection{Content Loss}
Content loss is introduced as a condition in an img2img translation GAN to directly evaluate how similar a generated image is to a real image. The loss consists of the pixel-wise L1 distance, the image gradient L1, and the image gradient structure losses. These losses are selected in different stages.

We designed two types of content losses suitable for the unsupervised and supervised learning process \eqref{eq2} as shown in Figure \ref{Figure 1}(B).

\begin{equation}
\begin{aligned}
\label{eq2}
\mathcal{L}_{ctn}=&\mathcal{L}_{ctn}^{{R\rightarrow M\rightarrow R}}+\mathcal{L}_{ctn}^{{M\rightarrow R\rightarrow M}}+\mathcal{L}_{ctn}^{{M\rightarrow R}}\\&+\mathcal{L}_{ctn}^{{R\rightarrow M}}   
\end{aligned}
\end{equation}

Where $\mathcal{L}_{ctn}^{{R\rightarrow M\rightarrow R}}$ and $\mathcal{L}_{ctn}^{{M\rightarrow R\rightarrow M}}$ are the RSI-OM-CyC losses, while $\mathcal{L}_{ctn}^{{M\rightarrow R}}$ and $\mathcal{L}_{ctn}^{{R\rightarrow M}}$ are the Direct-Content losses.

\textbf{RSI-OM-CyC Loss:}
When training on unpaired samples, the corresponding map is lacking when processing an unpaired remote sensing image, which is similar to that of an unpaired map. To compare the pixel values between generated and real images in this process, we employed cycle-consistency \cite{zhu_unpaired_2017} to build the unsupervised learning strategy. Cycle-consistency demands that the image translation cycle brings the input image back to the original image. The two "Remote Sensing Image - Styled Map Tiles Cycle Consistency (RSI-STM-CyC)" are in our task. Specifically, for each remote sensing image $x_R$, the following $x_R\rightarrow G_{R\rightarrow M}\left(x_R\right)\rightarrow\ G_{M\rightarrow R}\left(G_{R\rightarrow M}\left(x_R\right)\right)\approx\ x_R$ is presented as the RMR-Cycle. For each styled map $x_M$, the following $x_M\rightarrow\ G_{M\rightarrow R}\left(x_M\right)\rightarrow\ G_{R\rightarrow M}\left(G_{M\rightarrow R}\left(x_M\right)\right)\approx\ x_M$ is presented as the MRM-Cycle. For the RMR-Cycle, only the pixel-wise L1 loss ($\mathcal{L}_{L1}$) is employed to guide the model in order to improve the cycle-consistency by decreasing the mean difference in the pixel value between a generated remote sensing image and its ground truth \eqref{eq3}. By contrast, for the MRM-Cycle, the cycle-content loss is complicated. In addition to the pixel-wise L1 loss, the image gradient L1 loss (GraL1 loss, $\mathcal{L}_{gral1}$) and the image gradient structure loss (GraStr loss, $\mathcal{L}_{grastr}$) are employed \eqref{eq4} to guide the model in order to improve the cycle-consistency through improving topological consistency, which will be discussed in detail in Section 3.3.

\begin{equation}
\begin{aligned}
\label{eq3}
\mathcal{L}_{ctn}^{{R\rightarrow M\rightarrow R}}&=\lambda_{L1u}\mathcal{L}_{L1}\\&=\lambda_{L1u}\mathbb{E}_{x_R\sim p}[\|{G}_{M\rightarrow R}({G}_{R\rightarrow M}(x_R)-x_R\|_1]
\end{aligned}
\end{equation}
\begin{equation}
\begin{aligned}
\label{eq4}
&\mathcal{L}_{ctn}^{{M\rightarrow R\rightarrow M}}={\lambda_{L1u}\mathcal{L}_{L1}+\mathcal{L}}_{gral1}+\mathcal{L}_{grastr}\\&=\lambda_{L1u}\mathbb{E}_{x_M\sim p}[\|G_{R\rightarrow M}(G_{M\rightarrow R}(x_M))-x_M\|_1]\\&+\mathbb{E}_{x_M\sim p}[\|\mathcal{G}(G_{R\rightarrow M}(G_{M\rightarrow R}(x_M)))-\mathcal{G}(x_M)\|_1]\\&+
\mathbb{E}_{x_M\sim p}[2\\&-\frac{1}{N-1}\sum_{j=0}^{N-2}\frac{|\sigma_{\mathcal{G}_j(x_M)\mathcal{G}_j(G_{R\rightarrow M}(G_{M\rightarrow R}(x_M)))}|+C_1}{|\sigma_{\mathcal{G}_j(x_M)}\sigma_{\mathcal{G}_j(G_{R\rightarrow M}(G_{M\rightarrow R}(x_M)))}|+C_1}\\&-\frac{1}{M-1}\sum_{i=0}^{M-2}\frac{|\sigma_{\mathcal{G}_i(x_M)\mathcal{G}_i(G_{R\rightarrow M}(G_{M\rightarrow R}(x_M)))}|+C_2}{|\sigma_{\mathcal{G}_i(x_M)}\sigma_{\mathcal{G}_i(G_{R\rightarrow M}(G_{M\rightarrow R}(x_M)))}|+C_2}]
\end{aligned}
\end{equation}

Where $\lambda$ is a coefficient for fine-tuning; $\mathcal{L}_{L1}$ is the pixel-wise L1 loss; $\mathcal{L}_{gral1}$ is the image gradient L1 loss; and $\mathcal{L}_{grastr}$ is the image gradient structure loss.

\textbf{Direct-Content Loss:} Direct-Content Loss is introduced to generate remote sensing images and styled map tiles for supervised strategy. Direct-Content Loss has two forms for different generations. For generation from a styled map tile to a remote sensing image, only the pixel-wise L1 loss is employed to measure the mean difference in the pixels between a generated remote sensing image and its ground truth \eqref{eq5}. For the generation from a remote sensing image to a styled map tile, the GraL1 loss and the GraStr loss are employed to measure the topological consistency between the generated map and the ground truth, in addition to the pixel-wise L1 loss \eqref{eq6}.

\begin{equation}
    \begin{aligned}
    \label{eq5}
    \mathcal{L}_{ctn}^{M\rightarrow R}=\lambda_{L1}\mathcal{L}_{L1}=\lambda_{L1}\mathbb{E}_{x_M\sim p}[\|G_{M\rightarrow R}(x_M)-x_R\|_1]
    \end{aligned}
\end{equation}
\begin{equation}
    \begin{aligned}
    \label{eq6}
    &\mathcal{L}_{ctn}^{R\rightarrow M}=\lambda_{L1}\mathcal{L}_{L1}+\mathcal{L}_{gral1}+\mathcal{L}_{grastr}\\&=\lambda_{L1}\mathbb{E}_{x_R\sim p}[\|G_{R\rightarrow M}(x_R)-x_M\|_1]\\&+\mathbb{E}_{x_R\sim p}[\|\mathcal{G}(G_{R\rightarrow M}(x_R))-\mathcal{G}(x_M)\|_1]\\&+\mathbb{E}_{x_R\sim p}[2-\frac{1}{N-1}\sum_{j=0}^{N-2}\frac{|\sigma_{\mathcal{G}_j(x_M)\mathcal{G}_j(G_{R\rightarrow M}(x_R))}|+C_1}{\sigma_{\mathcal{G}_j(x_M)}\sigma_{\mathcal{G}_j(G_{R\rightarrow M}(x_R))}+C_1}\\&-\frac{1}{M-1}\sum_{i=0}^{M-2}\frac{|\sigma_{\mathcal{G}_i(x_M)\mathcal{G}_i(G_{R\rightarrow M}(x_R))}|+C_2}{\sigma_{\mathcal{G}_i(x_M)}\sigma_{\mathcal{G}_i(G_{R\rightarrow M}(x_R))}+C_2}]
    \end{aligned}
\end{equation}

Where $\lambda$ is a coefficient for fine tuning; $\mathcal{L}_{L1}$ is the pixel-wise L1 loss; $\mathcal{L}_{gral1}$ is the image gradient L1 loss; and $\mathcal{L}_{grastr}$ is the image gradient structure loss.

\subsubsection{Adversarial Loss}We employed the adversarial loss \cite{mirza_conditional_2014} in the model to guide the generator to learn high-frequency features, such as the style extracted by the discriminator,  \eqref{eq7}, \eqref{eq8}. In the unsupervised learning process, we only considered the adversarial loss corresponding to the first step generator and ignored the one corresponding to the second step. We therefore have the following:
\begin{equation}
    \begin{aligned}
    \label{eq7}
    \mathcal{L}_{adv}^{M\rightarrow R}&=\mathbb{E}_{x_R\sim p}[\log D_R(x_R)]\\&+\mathbb{E}_{x_M\sim p}[\log (1-D_R(G_{M\rightarrow R}(x_M)))]
    \end{aligned}
\end{equation}
\begin{equation}
    \begin{aligned}
    \label{eq8}
    \mathcal{L}_{adv}^{R\rightarrow M}&=\mathbb{E}_{x_M\sim p}[\log D_M(x_M)]\\&+\mathbb{E}_{x_R\sim p}[\log (1-D_M(G_{R\rightarrow M}(x_R)))]
    \end{aligned}
\end{equation}

\subsubsection{Identity Loss}$G_{R\rightarrow M}$ is a mapping from R to M. However, this mapping is a homomorphic mapping because many features that only belong to remote sensing images would be eliminated when a remote sensing image is translated to a map. We hope this mapping can be similar to an isomorphic mapping, whereby the network can output a very similar map when we input a map to $G_{R\rightarrow M}$. Thus, we introduced the identity loss derived from identity mapping \cite{taigman_unsupervised_2016} to improve the structure information mapping of the generator. We therefore have the following:

\begin{equation}
    \begin{aligned}
    \label{eq9}
    \mathcal{L}_{idt}^{R\rightarrow M}=\mathbb{E}_{x_M\sim p}[\|G_{R\rightarrow M}(x_M)-x_M\|_1]
    \end{aligned}
\end{equation}
\begin{equation}
    \begin{aligned}
    \label{eq10}
    \mathcal{L}_{idt}^{M\rightarrow R}=\mathbb{E}_{x_R\sim p}[\|G_{M\rightarrow R}(x_R)-x_R\|_1]
    \end{aligned}
\end{equation}

\subsection{Topological Consistency Loss}
In contrast to a general image translation task, the map tiles generation task must focus on the topological relationship, which is the structural information of objects on a map. However, the common L1 loss does not consider the topological relationship. Therefore we added a topological consistency loss as our optimization objective into the content loss in order to optimize the topological relationship of the generated map tiles, which consists of the image gradient L1 loss and the image structure loss as follows \eqref{eq11}:

\begin{equation}
    \begin{aligned}
    \label{eq11}
    &\mathcal{L}_{TopoCons}^{R\rightarrow M}=\mathcal{L}_{gral1}+\mathcal{L}_{grastr}\\&=\mathbb{E}_{x_R\sim p}[\|\mathcal{G}(G_{R\rightarrow M}(x_R))-\mathcal{G}(x_M)\|_1]\\&+\mathbb{E}_{x_R\sim p}[2-\frac{1}{N-1}\sum_{j=0}^{N-2}\frac{|\sigma_{\mathcal{G}_j(x_M)\mathcal{G}_j(G_{R\rightarrow M}(x_R))}|+C_1}{\sigma_{\mathcal{G}_j(x_M)}\sigma_{\mathcal{G}_j(G_{R\rightarrow M}(x_R))}+C_1}\\&-\frac{1}{M-1}\sum_{i=0}^{M-2}\frac{|\sigma_{\mathcal{G}_i(x_M)\mathcal{G}_i(G_{R\rightarrow M}(x_R))}|+C_2}{\sigma_{\mathcal{G}_i(x_M)}\sigma_{\mathcal{G}_i(G_{R\rightarrow M}(x_R))}+C_2}]
    \end{aligned}
\end{equation}

Where $\mathcal{L}_{gral1}$ is the image gradient L1 loss, and $\mathcal{L}_{grastr}$ is the image gradient structure loss.

Specifically, the topological consistency loss only works during the generation from a remote sensing image to a styled map, which includes the direct translation from a real remote sensing image to a styled map at the supervised stage and the translation from a generated remote sensing image to a styled map at the unsupervised stage in our model.

\subsubsection{Image Gradient L1 Loss}Objects on a map tile consist of points, lines, curves, and areas, and the topological relationship of objects can be presented by the topological relationship of the pixels in their outline as well as at the edge. The edge curves of objects divide a map tile into different areas for various objects. Thus, we can transform the topological relationship of different objects into the topological relationship of their edge curves.

\begin{figure}[ht]
\centering
\includegraphics[scale=0.5]{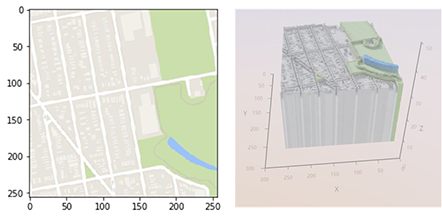}
\caption{A 256$\times$256 Google map tile and its 3d map, in which the z-axis is the luminance.}
\label{Fig2}
\end{figure}

Hence, we extract the edge curves of objects on a map tile. The gradient is often used to extract the edge of an image in image processing \cite{canny_computational_1986}. Figure \ref{Fig2} shows the generated 3D image of a Google map tile, where the x-axis and y-axis are the length and width of the image, respectively, and the z-axis is the luminance of every pixel. An object with the same color is an isohypsic surface, and its edge consists of points, of which the gradient value is large. Thus, we extract the gradient image of a map as its edge feature.

For a 256$\times$256 image, we set the pixel value at the point ${(i, j)}$ as ${f(i, j)}$. Then the row gradient of this point is:
\begin{equation}
\label{eq12}
    g_x(i, j)=f(i, j+1)-f(i,j)
\end{equation}
and the column gradient is:
\begin{equation}
\label{eq13}
   g_y(i, j)=f(i+1, j)-f(i,j) 
\end{equation}
Hence the gradient of this point is:
\begin{equation}
\label{eq14}
    g(i, j)=\sqrt[2]{g_x(i,j)^2+g_y(i,j)^2}
\end{equation}
We can then obtain a 255$\times$255 matrix $\mathcal{G}$ consisting of $\emph{g(i,j)}$ as the gradient map of the image.

\begin{figure}[ht]
\centering
\includegraphics[scale=0.5]{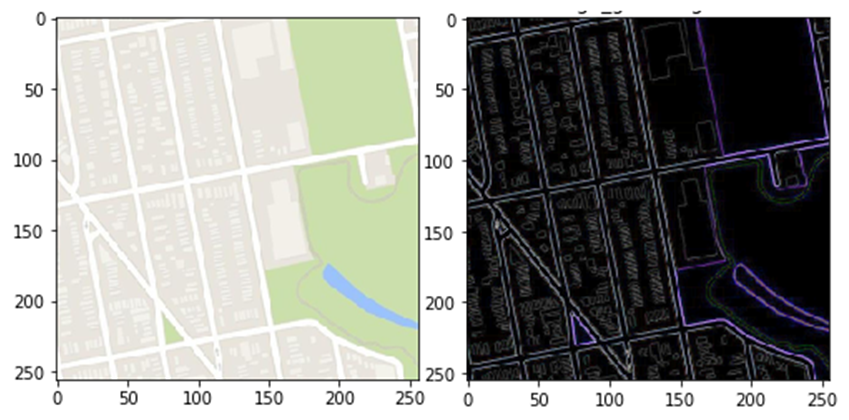}
\caption{A 256$\times$256 Google map tile and its 255$\times$255 gradient map.}
\label{Figure 3}
\end{figure}

Figure \ref{Figure 3} presents a Google map tile and its gradient image. The gradient image shows a basic outline of all objects and the outlines divide the gradient image into different areas. Each area is mapped to an object on the map, and each color lump of an object on the map is colored differently, whereas it is black inside its corresponding area on the gradient image (all pixel values are 0). Hence, the color does not change inside an area but changes only along the edge. Thus, a gradient map can preserve the topological relationships of objects and the differences among adjoining objects.

According to the gradient image, we proposed the image gradient L1 loss as follows:

\begin{equation}
\label{eq15}
    \mathcal{L}_{gral1}=\mathbb{E}_{x_R\sim p}[\|\mathcal{G}(G_{R\rightarrow M}(x_R))-\mathcal{G}(x_M)\|_1]
\end{equation}

Where $\mathcal{G}(x_M)$ is the gradient image of the real map $x_M$ and $\mathcal{G}(G_{R\rightarrow M}(x_R))$is the gradient image of the generated map $G_{R\rightarrow M}\left(x_R\right)$. By this loss, we hope the generator decreases the L1 distance between the image gradient maps of generated maps and their corresponding ground truths. By decreasing this distance, the global gradient as well as the pixel increment among different objects of generated maps can be drawn close to the ground truths.

\subsubsection{Image Gradient Structure Loss}However, we have not tackled the topological relationship consistency between the generated and real maps. As mentioned before, the topological relationship of objects can be transformed into the the topological relationship of the edge curves of the objects. Furthermore, the topological relationship of the edge curves can be transformed into the topological relationship among each point on the edge curves of the objects, as well as the topological relationship among each point on the gradient image.

For $255\times255$ gradient images of real and generated maps, their pixel matrices $\mathcal{G}(x_M)$ and $\mathcal{G}\left(G_{R\rightarrow M}\left(x_R\right)\right)$ have 255 rows and 255 columns. For the ${jth}$ column (${ith}$ row), the pixel values of points on this column (row) are comprised of an ${i}$-dimension (${j}$-dimension) random variable $\mathcal{G}_j (\mathcal{G}_i)$. For the same column (row) of the generated image and the real map, we hope they are correlative. Due to this possibility, the random variable $\mathcal{G}_j (\mathcal{G}_i)$ preserves the topological relationship of the points in $\mathcal{G}_j (\mathcal{G}_i)$ by its dimensions, and $\mathcal{G}_j(x_M)$ and $\mathcal{G}_j\left(G_{R\rightarrow M}\left(x_R\right)\right)$ are very topologically consistent when they are highly correlative.

We employ the Pearson correlation coefficient to measure the correlation between $\mathcal{G}_j(x_M)$ and $\mathcal{G}_j\left(G_{R\rightarrow M}\left(x_R\right)\right)$, which is as follows:

\begin{equation}
    \begin{aligned}
    \label{eq16}
    \rho(\mathcal{G}_j(x_M),\mathcal{G}_j(G_{R\rightarrow M}(x_R)))=\frac{\sigma_{\mathcal{G}_j(x_M)\mathcal{G}_j(G_{R\rightarrow M}(x_R))}}{\sigma_{\mathcal{G}_j(x_M)}\sigma_{\mathcal{G}_j(G_{R\rightarrow M}(x_R))}}
    \end{aligned}
\end{equation}

Where $\sigma_{\mathcal{G}_j(x_M)\mathcal{G}_j\left(G_{R\rightarrow M}\left(x_R\right)\right)}$ is the covariance of $\mathcal{G}_j(x_M)$ and $\mathcal{G}_j\left(G_{R\rightarrow M}\left(x_R\right)\right)$; $\sigma_{\mathcal{G}_j(x_M)}$ is the standard deviation of $\mathcal{G}_j(x_M)$; $\sigma_{\mathcal{G}_j\left(G_{R\rightarrow M}\left(x_R\right)\right)}$ is the standard deviation of $\mathcal{G}_j\left(G_{R\rightarrow M}\left(x_R\right)\right)$; and $\rho(\mathcal{G}_j(x_M),\mathcal{G}_j(G_{R\rightarrow M}(x_R)))\in[-1,1]$. When $\rho$ is closer to 1, $\mathcal{G}_j(x_M\ )$ and $\mathcal{G}_j\left(G_{R\rightarrow M}\left(x_R\right)\right)$ are more correlative.

Due to the input image having an M$\times$N scale with N columns, we calculate the mean value of all $\rho(\mathcal{G}_j(x_M),\mathcal{G}_j(G_{R\rightarrow M}(x_R)))$:

\begin{equation}
    \begin{aligned}
    \label{eq17}
    \rho(\mathcal{G}_j(x_M),\mathcal{G}_j(G_{R\rightarrow M}&(x_R)))_{mean}=\\&\frac{1}{N-1}\sum_{j=0}^{N-2}\frac{\sigma_{\mathcal{G}_j(x_M)\mathcal{G}_j(G_{R\rightarrow M}(x_R))}}{\sigma_{\mathcal{G}_j(x_M)}\sigma_{\mathcal{G}_j(G_{R\rightarrow M}(x_R))}}
    \end{aligned}
\end{equation}

We have:

\begin{equation}
    \begin{aligned}
    \label{eq18}
    \rho(\mathcal{G}_i(x_M),\mathcal{G}_i(G_{R\rightarrow M}&(x_R)))_{mean}=\\&\frac{1}{M-1}\sum_{i=0}^{M-2}\frac{\sigma_{\mathcal{G}_i(x_M)\mathcal{G}_i(G_{R\rightarrow M}(x_R))}}{\sigma_{\mathcal{G}_i(x_M)}\sigma_{\mathcal{G}_i(G_{R\rightarrow M}(x_R))}}
    \end{aligned}
\end{equation}

for all rows. Thus, we proposed the image gradient structure loss:

\begin{equation}
    \begin{aligned}
    \label{19}
    \mathcal{L}_{grastr}=&\mathbb{E}_{x_R\sim p}[2\\&-\frac{1}{N-1}\sum_{j=0}^{N-2}\frac{|\sigma_{\mathcal{G}_j(x_M)\mathcal{G}_j(G_{R\rightarrow M}(x_R))}|+C_1}{\sigma_{\mathcal{G}_j(x_M)}\sigma_{\mathcal{G}_j(G_{R\rightarrow M}(x_R))}+C_1}\\&-\frac{1}{M-1}\sum_{i=0}^{M-2}\frac{|\sigma_{\mathcal{G}_i(x_M)\mathcal{G}_i(G_{R\rightarrow M}(x_R))}|+C_2}{\sigma_{\mathcal{G}_i(x_M)}\sigma_{\mathcal{G}_i(G_{R\rightarrow M}(x_R))}+C_2}]
    \end{aligned}
\end{equation}

Where $C_1$ and $C_2$ are constant to maintain a stable loss when $\left|\sigma_{\mathcal{G}_j\left(x_M\ \right)}\sigma_{\mathcal{G}_j\left(G_{R\rightarrow M}\left(x_R\right)\right)}\right|$ or $\left|\sigma_{\mathcal{G}_i(x_M\ )}\sigma_{\mathcal{G}_i\left(G_{R\rightarrow M}\left(x_R\right)\right)}\right|$ is very close to 0. When this loss is decreasing, the correlations of each column and each row of the gradient images of the generated map and the real map are increasing. Consequently, the generator can improve the topological consistency between the generated and real maps.

\subsection{Model Implements}
\subsubsection{Network Architecture}Inspired by the design in \cite{zhu_unpaired_2017}, SMAPGAN builds its generators with a 9-ResNet-Blocks architecture and PatchGAN's \cite{zhu_unpaired_2017,demir_patch-based_2018} discriminators with a size of 70, which show an impressive performance at img2img translation. Specifically, each generator network consists of two downsampling convolutional layers with 2-stride, nine residual blocks \cite{he_deep_2016}, and two subsequent transposed convolutional layers of similar strides for upsampling. Instance normalization \cite{taigman_unsupervised_2016} is leveraged in the generator, but not in the discriminator.

\subsubsection{Avoid Steganography During Unsupervised Stage} \cite{chu_cyclegan_2017} revealed that the CycleGAN model tends to add high-frequency noise to hide information in intermediate results in order to achieve improved reconstruction. To avoid unnecessary noise, we set up a trick during our training. Specifically, we applied cycle-consistency to extend the availability of the unpaired training, which causes the models to learn steganography. Although this feature helps the generator recover the original image in the work based on cycle-consistency, steganography brings noise that interrupts the correct structure feature of the maps in our map-generated task. To avoid this issue, we freeze the weight of the first step generator in the two-step cycle reconstruction (which tends to learn steganography) when training in the unsupervised stage. This rule will be activated after several epochs (we denote the threshold as $t_s$), which allows for cycle-consistency to help accelerate the model convergence early on.

\subsection{Topological Consistency Measurement: ESSI}
\subsubsection{ESSI: Edge Structural Similarity Index}
A traditional image quality assessment, including MSE and SSIM, can be introduced to assess the quality of the generated maps. MSE can measure the global pixel-wise similarity between the generated maps and their corresponding ground truth, whereas the SSIM can measure the similarity of the luminance, contrast, and structure between generated maps and their ground truth. However, besides MSE and SSIM, we hope to employ a suitable metric for measuring the topological consistency between a generated map and a real map. In Section 3.3, we discussed that the topological relationship of objects on maps can be preserved by the edge curves of objects. Based on this situation, we proposed an edge structural similarity index (ESSI):

\begin{equation}
    \begin{aligned}
    \label{eq20}
    ESSI(&G(x_R),x_M)=\\&\frac{(|\sigma_{\mathcal{E}(G(x_R))\mathcal{E}(x_M)}|+C_1)(2\mu_{\mathcal{E}(G(x_R))}\mu_{\mathcal{E}(x_M)}+C_2)}{(\sigma_{\mathcal{E}(G(x_R))}\sigma_{\mathcal{E}(x_M)}+C_1)(\mu^2_{\mathcal{E}(G(x_R))}+\mu^2_{\mathcal{E}(x_M)}+C_2)}
    \end{aligned}
\end{equation}

Where $\mathcal{E}\left(G\left(x_R\right)\right)$ and $\mathcal{E}\left(x_M\right)$ are the pixel matrix of the edge images of a generated styled map tiles and its ground truth; $\sigma_{\mathcal{E}\left(G\left(x_R\right)\right)}$ and $\sigma_{\mathcal{E}\left(x_M\right)}$ are the standard deviations of $\mathcal{E}\left(G\left(x_R\right)\right)$ and $\mathcal{E}\left(x_M\right)$, respectively; $\sigma_{\mathcal{E}\left(G\left(x_R\right)\right)\mathcal{E}\left(x_M\right)}$ is the covariance between $\mathcal{E}\left(G\left(x_R\right)\right) and \mathcal{E}\left(x_M\right)$; $C_1$ is a constant to keep ESSI stable when $\sigma_{\mathcal{E}\left(G\left(x_R\right)\right)}\sigma_{\mathcal{E}\left(x_M\right)}$ is close to 0; $\mu_{\mathcal{E}\left(G\left(x_R\right)\right)}$ and $\mu_{\mathcal{E}\left(x_M\right)}$ are the mean values of $\mathcal{E}\left(G\left(x_R\right)\right)$ and $\mathcal{E}\left(x_M\right)$, respectively; and $C_2$ is a constant to keep the metric stable when ${\mu_{\mathcal{E}\left(G\left(x_R\right)\right)}}^2{{+\mu}_{\mathcal{E}\left(x_M\right)}}^2$ is close to 0.

Detailed proof of the ESSI is presented in the appendix.

\begin{figure}[ht]
\centering
\includegraphics[scale=0.6]{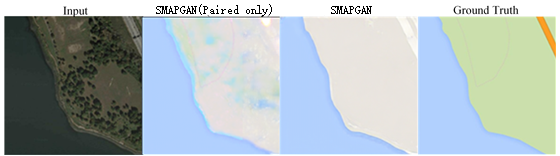}
\caption{Input remote sensing image, two generated map tiles using different models and the corresponding Google map tile (ground truth). The output of SMAPGAN (Paired only) has an MSE of 148.50, an SSIM of 0.9531 and an ESSI of 0.3036. By contrast, the generated map tile of SMAPGAN has an MSE of 169.41, an SSIM of 0.9482 and an ESSI of 0.4299. Although the second map tile outperformed the third for the MSE and the SSIM, the third map tile has a higher ESSI and a better topological relationship of objects from a human visual perspective.}
\label{Figure 5}
\end{figure}

\subsubsection{Advantage of ESSI Compared to MSE and SSIM}
ESSI outperforms MSE and SSIM in terms of measuring topological consistency. Figure \ref{Figure 5} presents two map tiles generated from the same remote sensing image using different models. The map tile generated by SMAPGAN (paired only, which means it is only trained on paired training samples) obtains an MSE score of 148.50 and SSIM score of 0.9531, which are better than the output of SMAPGAN (trained on all training samples) with an MSE score of 169.41 and an SSIM score of 0.9482. From the MSE and SSIM scores, the second image has a better quality than the third. However, from the human visual perspective, the third image has a better topological relationship of objects than the second image. The third image presented a clear road on the top right corner and no blur on the grass. Thus, the ESSI fits the human visual perspective more than the MSE and SSIM when evaluating the topological consistency of the two images (the ESSI of the second image is 0.3036 while the third is 0.4299).

The calculation method of the MSE and SSIM results in this phenomenon. The blurred part of the second map has a high approximate pixel value with the ground truth, which increases the MSE and SSIM. However, these blurred parts only decreased the ESSI due to a fault of the topological relationship, which resulted in a blur.

\begin{figure}[ht]
\centering
\includegraphics[scale=0.6]{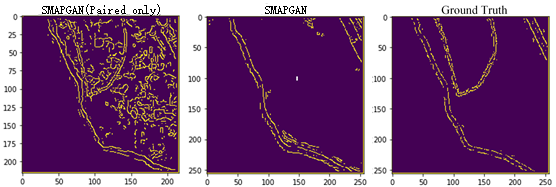}
\caption{Edge images of the generated maps of different models and the ground truth. The edge images extract the topological relationships more precisely with a low blur influence.}
\label{Figure 6}
\end{figure}
Figure \ref{Figure 6} shows the edge images for the output of the SMAPGAN (paired only) and SMAPGAN, as well as their corresponding ground truths. The blurred parts of the generated map of SMAPGAN (paired only) led to several more edge curves than the ground truth, which resulted in wrong topological consistency. The calculation of ESSI is based on the edge image, which seizes the topological relationship of objects more accurately with a lower blur influence. Thus, the topological consistency between the generated map tile and its ground truth can be evaluated precisely by the ESSI.
\section{Experiments and Results}
\subsection{Datasets and Baseline}
\subsubsection{Datasets}
We conducted our experiments on two datasets related to styled map tiles generating:

\textbf{Aerial photograph $\leftrightarrow$ Maps} \cite{isola_image--image_2017}: This dataset consists of 2194 aerial images and their corresponding styled map tiles scraped from Google Maps. All samples involve places in and around New York City. The aerial images are in the visible wavelengths and have a spatial resolution of 2.15 meters per pixel. This dataset is denoted as dataset A in the results tables.

\textbf{Simple styled maps $\leftrightarrow$ Target styled maps} \cite{kang_transferring_2019}: This dataset consists of approximately 1,088 simple styled map tiles at zoom 15 generated from Open Street Map (OSM) vector data, as well as 1,088 target styled map tiles matching the simple styled ones. In this dataset, the simple styled maps are used as the input. As opposed to rendering from vector data, the inputs are also images. We utilized this dataset to conduct experiments to check the applicability of the model among all styled map tiles generating tasks. This dataset is denoted as dataset B in the results tables.

All of these datasets were shuffled and randomly separated into paired and unpaired parts by different paired-ratios, and then were divided into training sets and testing sets. Specifically, we set the paired-ratios as 10\% and 50\% as a classical ratio for semi-supervised learning \cite{} to check how our model performed at different semi-supervised levels. The final datasets are denoted as A-10\%, A-50\%, B-10\%, and B-50\%. A-10\% and A-50\% share the same testing set from dataset A, and B-10\% and B-50\% share the same testing set from dataset B. In our training set, each remote sensing image has a corresponding map tile in the paired part. None of the remote sensing images or any maps have their corresponding images from the opposed domain in the unpaired part of our datasets.

\begin{figure}[ht]
\centering
\includegraphics[scale=0.55]{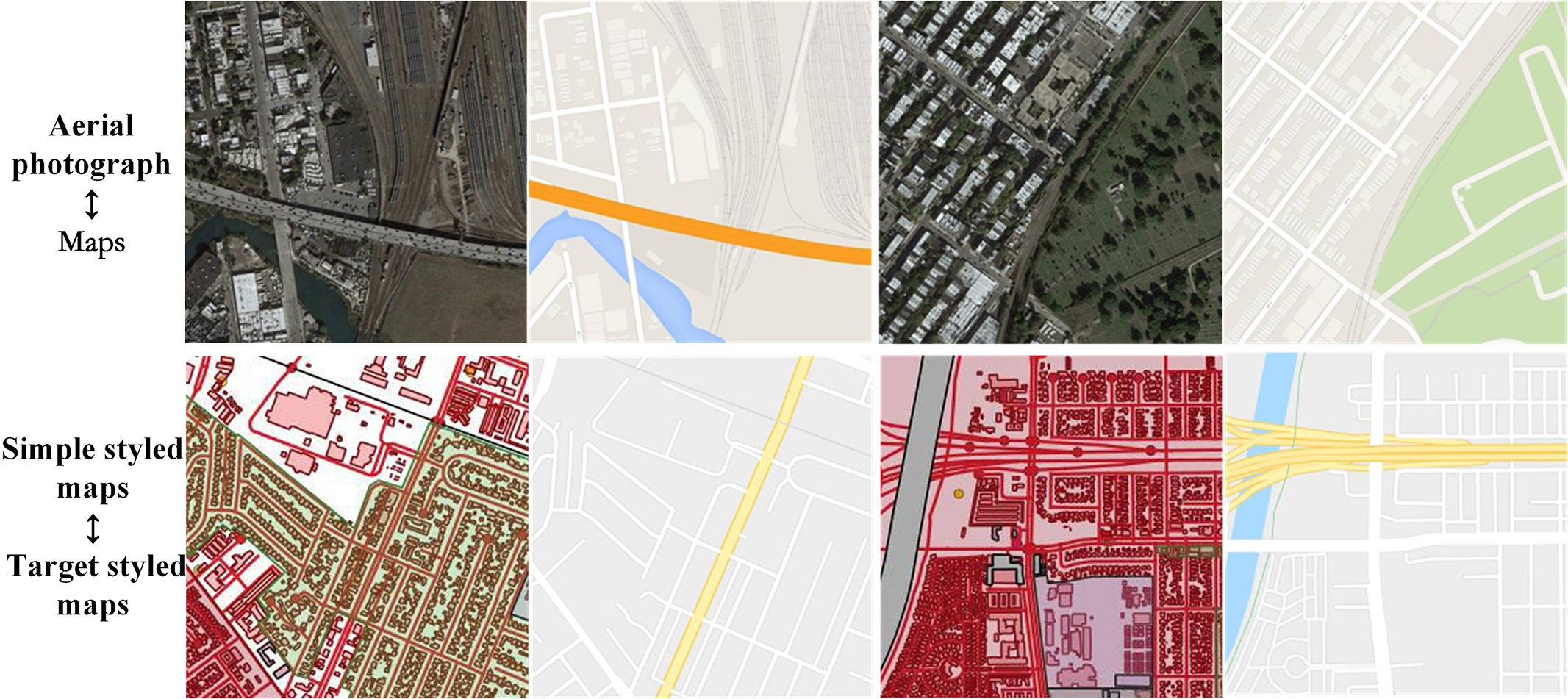}
\centering
\caption{Samples of the two types of dataset.}
\label{Figure 7}
\end{figure}

\subsubsection{Baseline}
We compared SMAPGAN with the five SOTA methods mentioned in Section 2, namely, Pix2Pix \cite{isola_image--image_2017}, CycleGAN \cite{zhu_unpaired_2017}, UNIT \cite{liu_unsupervised_2017}, AttentionCycleGAN \cite{alami_mejjati_unsupervised_2018} and GeoGAN \cite{ganguli_geogan:_2019}. Due to the limitation where Pix2Pix and GeoGAN could only be trained in a supervised manner, we ran them on the paired part of our dataset to see their performance and test the effectiveness of our topological consistency loss function. To perform a comparison, we trained SMAPGAN only on the same paired part. For other models, we fed all the samples to them in their acceptable methods.

\subsection{Metrics}
\begin{figure*}[htbp]
\centering
\includegraphics[scale=0.55]{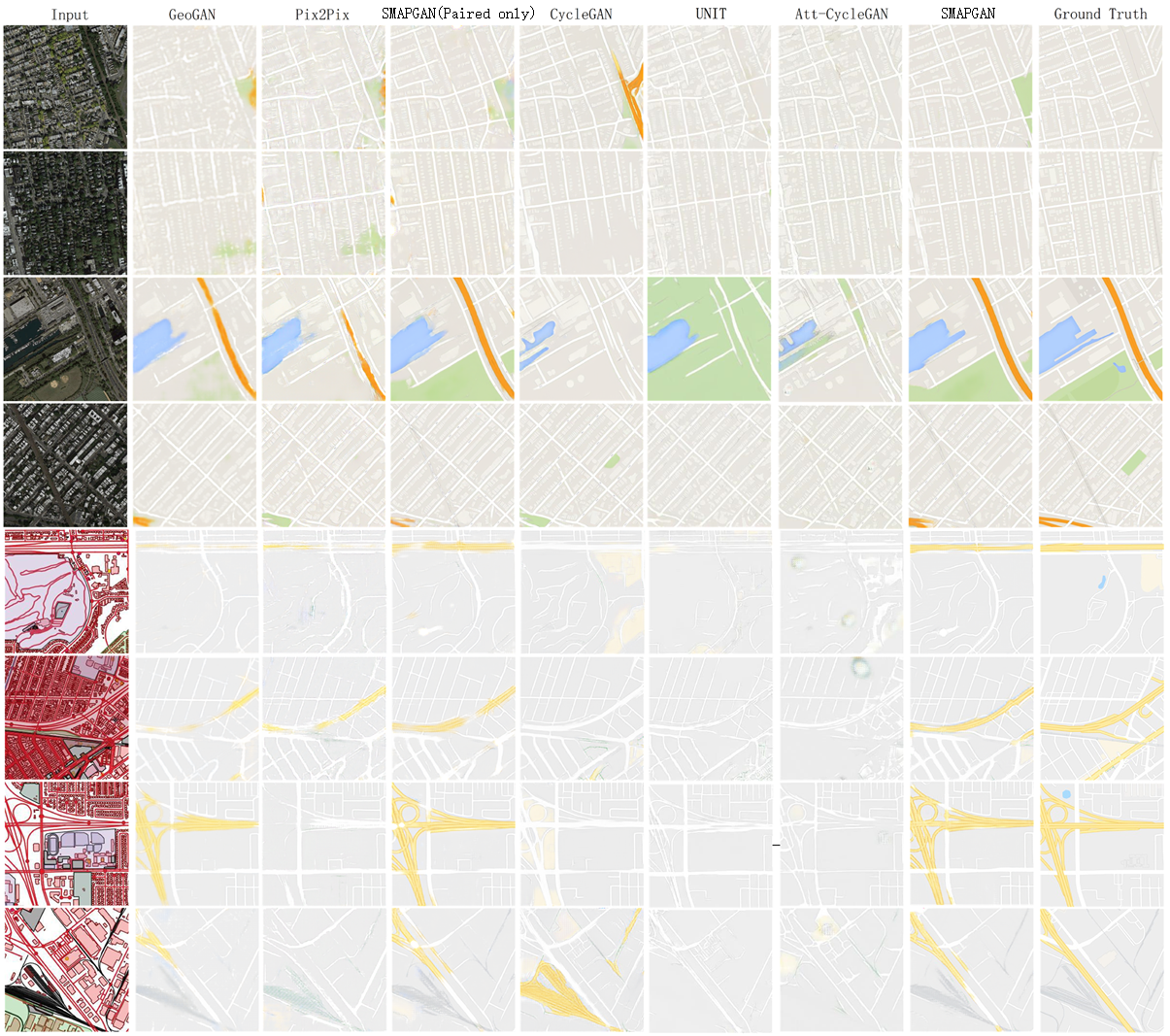}
\caption{Some selected samples of each dataset in the comparison experiment. Every two rows correspond with one of the four datasets listed in 4.1. Each row includes the input image, maps generated by five baselines and SMAPGAN, and the ground truth. Specifically, the maps generated by SMAPGAN trained on only the paired data and all data are presented respectively in each row.}
\label{Figure 8}
\end{figure*}
\begin{figure*}[htbp]
\centering
\includegraphics[scale=0.63]{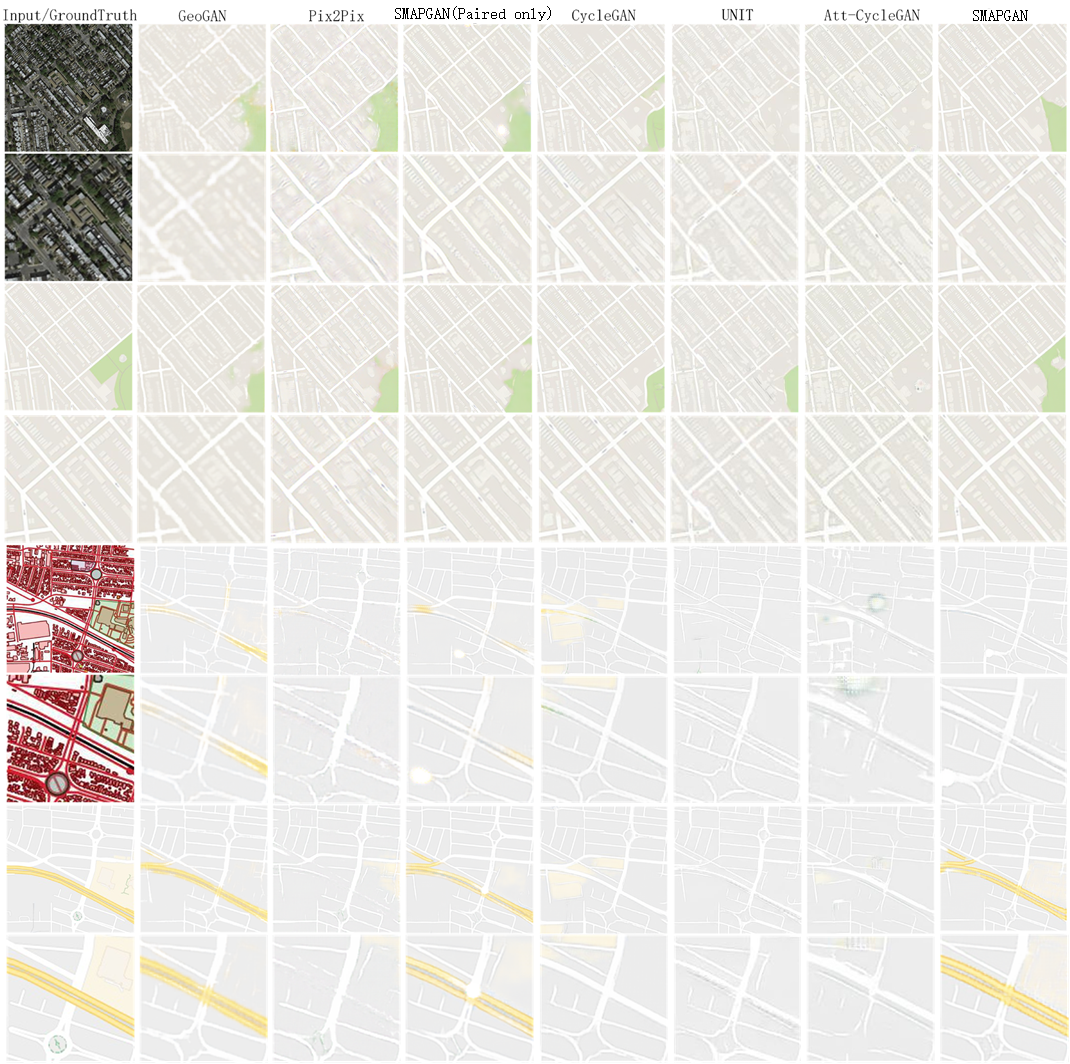}
\caption{Selected samples of each source dataset in a comparison experiment. Every four rows correspond with one of two source datasets. The first column of each four rows consists of the full-size input image, the magnified area (showing the structure feature) of the input image, the full-size ground truth, and the magnified area of the ground truth. The first two rows of other columns list the maps and their magnified area generated by each model trained on a 10\% paired-ratio training set and the two rows below list the maps and their magnified area generated by each model trained on a 50\% paired-ratio training set.}
\label{Figure 9}
\end{figure*}

We employed objective and subjective metrics to measure the quality of the generated map tiles. As each generated map tile has a corresponding ground truth, we applied a traditional full-reference image quality assessment, including MSE and SSIM. Also, the new full-reference image quality assessment ESSI proposed by us was employed to measure the topological consistency between the output map tile and its ground truth.

\subsubsection{Objective Metrics}
\textbf{MSE:} We used MSE as the first metric to measure the output images because it is commonly used in the image processing task. MSE is as follows:
\begin{equation}
    \begin{aligned}
    \label{eq25}
    MSE(&G(x_R),x_M)=\\&\frac{1}{M\times N}\sum_{i=0}^{M-1}\sum_{j=0}^{N-1}(p_{G(x_R)}(i,j)-p_{x_M}(i,j))^2
    \end{aligned}
\end{equation}

Where $M$ and $N$ are the length and width of the image, respectively; and $p_{G\left(x_R\right)}\left(i,j\right)$ and $p_{x_m}\left(i,j\right)$ are the pixel values at the pixel $(i, j)$ of the generated map $G(x_R)$ and the real map $x_M$, respectively.

\textbf{SSIM}: We used the SSIM to measure the structural similarity between the generated map and its ground truth. The SSIM is as follows:
\begin{equation}
    \begin{aligned}
    \label{eq26}
    SSIM(&G(x_R),x_M)=\\&\frac{(2\mu_{G(x_R)}\mu_{x_M}+C_1)(2\sigma_{G(x_R)x_M}+C_2)}{(\mu_{G(x_R)}^2+\mu_{x_M}^2+C_1)(\sigma_{G(x_R)}^2+\sigma_{x_M}^2+c_2)}
    \end{aligned}
\end{equation}

Where $\mu_{G\left(x_R\right)}$ and $\mu_{x_M}$ are the means of $G\left(x_R\right)$ and $x_m$, respectively; ${\sigma_{G\left(x_R\right)}}^2$ and ${\sigma_{x_M}}^2$ are the variances of $G\left(x_R\right)$ and $x_M$, respectively; $\sigma_{G\left(x_R\right)x_M}$ is the covariance between $G\left(x_R\right)$ and $x_M$; and $C_1$ and $C_2$ are constants to keep the metric stable when ${\mu_{G\left(x_R\right)}}^2+{\mu_{x_m}}^2$ and ${\sigma_{G\left(x_R\right)}}^2{{+\sigma}_{x_m}}^2$ are close to 0.

\textbf{ESSI:} We used the ESSI \eqref{eq20} proposed in Section 3.5 as our metric as well.

\subsubsection{Subjective Metric}
Since objective metrics could not completely fit the human visual system, we designed a human perceptual test to check if our model generates more valid styled map tiles than baselines. We randomly picked participants from graduates majoring in GIS and built up a web application as a platform to conduct the quality assessment for all the models on the four datasets, which contained 290, 290, 515, and 515 samples. The platform offered the participants a website that presents a multiple-choice question containing eight pictures randomly picked from one of the four datasets for each time, namely, the inputted remote sensing image, ground truth corresponding to the input, and styled maps generated from the input by the six models. These six models included five baselines and our SMAPGAN trained with semi-supervised setup on the full data. The participants should select the relatively the best result that could best describe the input and is most similar to the ground truth among selections. Meanwhile, the system would count the frequency that every model is selected. Finally, the statistics revealed the performance of every model.

\begin{table*}[htbp]
  \centering
  \caption{Training and Testing Time on Different Datasets}
    \begin{tabular}{p{6em}cccc}
    \toprule
          & \textbf{amount of training samples } & \textbf{amount of testing samples} & \textbf{training time (h:m:s)} & \textbf{testing time (h:m:s)} \\
    \midrule
    \textbf{A-10\%} & 1768  & 1030  & 12:18:24 & 0:02:45 \\
    \textbf{A-50\%} & 2240  & 1030  & 14:36:30 & 0:03:04 \\
    \textbf{B-10\%} & 840   & 580   & 5:53:38 & 0:01:21 \\
    \textbf{B-50\%} & 1064  & 580   & 6:51:41 & 0:01:29 \\
    \bottomrule
    \end{tabular}%
  \label{tab:addlabel}%
\end{table*}%

\begin{table*}[htbp]
  \centering
  \caption{MSE (the smaller the better)}
    \begin{tabular}{p{7.5em}cccccccc}
    \toprule
    \multirow{2}[4]{*}{\textbf{Model\textbackslash{}Dataset}} & \multicolumn{2}{p{8.38em}<{\centering}}{\textbf{A-10\%}} & \multicolumn{2}{p{8.38em}<{\centering}}{\textbf{A-50\%}} & \multicolumn{2}{p{8.38em}<{\centering}}{\textbf{B-10\%}} & \multicolumn{2}{p{8.38em}<{\centering}}{\textbf{B-50\%}} \\
\cmidrule{2-9}    \multicolumn{1}{c}{} & \multicolumn{1}{p{4.19em}}{RGBmean} & \multicolumn{1}{p{4.19em}}{Luminance} & \multicolumn{1}{p{4.19em}}{RGBmean} & \multicolumn{1}{p{4.19em}}{Luminance} & \multicolumn{1}{p{4.19em}}{RGBmean} & \multicolumn{1}{p{4.19em}}{Luminance} & \multicolumn{1}{p{4.19em}}{RGBmean} & \multicolumn{1}{p{4.19em}}{Luminance} \\
    \midrule
    GeoGAN & 449.9124 & 275.8598 & 193.0430 & 129.7763 & 91.8676 & 53.7895 & 70.8694 & 42.5808 \\
    Pix2Pix & 382.4626 & 253.9473 & 296.8432 & 205.6530 & 111.4474 & 68.0716 & 116.1450 & 67.0667 \\
   SMAPGAN(paired) & 323.7949 & 199.8037 & 174.6797 & 116.9807 & 94.2073 & 55.3836 &71.5514  & 43.8748 \\
    CycleGAN &449.9124 & 275.8598 & 439.5733 & 265.6958 & 173.4133 & 110.6720 & 167.6997 & 104.2759 \\
    UNIT  & 525.7523 & 305.2930 &483.8580 & 277.7960 & 142.5751 & 90.8529 & 127.1820 & 71.7548 \\
    Att-CycleGAN & 524.8167 & 299.4017 & 501.5179 & 291.2331 & 165.3915 & 114.7556 & 148.6484 & 98.4731 \\
    SMAPGAN & \textit{\textbf{385.5125}} & \textit{\textbf{226.1506}} & \textit{\textbf{176.3012}} & \textit{\textbf{116.7916}} & \textit{\textbf{100.4296}} & \textit{\textbf{58.0107}} & \textit{\textbf{75.2226}} & \textit{\textbf{44.3301}} \\
    \bottomrule
    \end{tabular}%
  \label{tab:addlabel}%
\end{table*}%
\begin{table*}[htbp]
  \centering
  \caption{SSIM (the closer to 1 the better)}
    \begin{tabular}{p{7.5em}cccccccc}
    \toprule
    \multirow{2}[4]{*}{\textbf{Model\textbackslash{}Dataset}} & \multicolumn{2}{p{8.38em}<{\centering}}{\textbf{A-10\%}} & \multicolumn{2}{p{8.38em}<{\centering}}{\textbf{A-50\%}} & \multicolumn{2}{p{8.38em}<{\centering}}{\textbf{B-10\%}} & \multicolumn{2}{p{8.38em}<{\centering}}{\textbf{B-50\%}} \\
\cmidrule{2-9}    \multicolumn{1}{c}{} & \multicolumn{1}{p{4.19em}}{RGBmean} & \multicolumn{1}{p{4.19em}}{Luminance} & \multicolumn{1}{p{4.19em}}{RGBmean} & \multicolumn{1}{p{4.19em}}{Luminance} & \multicolumn{1}{p{4.19em}}{RGBmean} & \multicolumn{1}{p{4.19em}}{Luminance} & \multicolumn{1}{p{4.19em}}{RGBmean} & \multicolumn{1}{p{4.19em}}{Luminance} \\
    \midrule
    GeoGAN & 0.7363 & 0.7534 & 0.7889 & 0.8008 & 0.8408 & 0.8483 & 0.8606 & 0.8667 \\
    Pix2Pix & 0.6657 & 0.6862 & 0.6998 & 0.7243 & 0.8029 & 0.8108 & 0.8249 & 0.8318 \\
   SMAPGAN(paired) & 0.7419 & 0.7572 & 0.7915 & 0.8029 & 0.8432 & 0.8506 & 0.8644 & 0.8698 \\
    CycleGAN & 0.7140 & 0.7273 & 0.7166 & 0.7287 & 0.6802 & 0.6880 & 0.6893 & 0.6968 \\
    UNIT  & 0.6390 & 0.6534 & 0.7003 & 0.7155 & 0.7160 & 0.7217 & 0.8039 & 0.8105 \\
    Att-CycleGAN & 0.6568 & 0.6709 & 0.6708 & 0.6836 & 0.6270 & 0.6318 & 0.6710 & 0.6760 \\
    SMAPGAN & \textit{\textbf{0.7515}} & \textit{\textbf{0.7651}} & \textit{\textbf{0.7993}} & \textit{\textbf{0.8105}} & \textit{\textbf{0.8443}} & \textit{\textbf{0.8511}} & \textit{\textbf{0.8665}} & \textit{\textbf{0.8720}} \\
    \bottomrule
    \end{tabular}%
  \label{tab:addlabel}%
\end{table*}%
\begin{table*}[htbp]
  \centering
  \caption{ESSI (the closer to 1 the better)}
    \begin{tabular}{p{7.5em}cccccccc}
    \toprule
    \multirow{2}[4]{*}{\textbf{Model\textbackslash{}Dataset}} & \multicolumn{2}{p{8.38em}<{\centering}}{\textbf{A-10\%}} & \multicolumn{2}{p{8.38em}<{\centering}}{\textbf{A-50\%}} & \multicolumn{2}{p{8.38em}<{\centering}}{\textbf{B-10\%}} & \multicolumn{2}{p{8.38em}<{\centering}}{\textbf{B-50\%}} \\
\cmidrule{2-9}    \multicolumn{1}{c}{} & \multicolumn{1}{p{4.19em}}{RGBmean} & \multicolumn{1}{p{4.19em}}{Luminance} & \multicolumn{1}{p{4.19em}}{RGBmean} & \multicolumn{1}{p{4.19em}}{Luminance} & \multicolumn{1}{p{4.19em}}{RGBmean} & \multicolumn{1}{p{4.19em}}{Luminance} & \multicolumn{1}{p{4.19em}}{RGBmean} & \multicolumn{1}{p{4.19em}}{Luminance} \\
    \midrule
    GeoGAN & 0.1758 & 0.186 & 0.2541 & 0.2661 & 0.3961 & 0.3995 & 0.4464 & 0.4502 \\
    Pix2Pix & 0.1370 & 0.1475 & 0.1879 & 0.2047 & 0.3229 & 0.3277 & 0.3617 & 0.3673 \\
   SMAPGAN(paired) & 0.2162 & 0.2278 & 0.2916 & 0.3071 & 0.4076 & 0.4124 & 0.4557 & 0.4605 \\
    CycleGAN & 0.2025 & 0.2179 & 0.2084 & 0.2215 & 0.0150 & 0.0185 & 0.0253 & 0.0294 \\
    UNIT & 0.0881 & 0.0939 & 0.1703 & 0.1787 & 0.1572 & 0.1594 & 0.3196 & 0.3199 \\
    Att-CycleGAN & 0.1418 & 0.1509 & 0.0881 & 0.1660 & 0.0730 & 0.0725 & 0.0309 & 0.0328 \\
    SMAPGAN & \textit{\textbf{0.2390}} & \textit{\textbf{0.2553}} & \textit{\textbf{0.3043}} & \textit{\textbf{0.3234}} & \textit{\textbf{0.4270}} & \textit{\textbf{0.4289}} & \textit{\textbf{0.4600}} & \textit{\textbf{0.4652}} \\
    \bottomrule
    \end{tabular}%
  \label{tab:addlabel}%
\end{table*}%
\begin{table*}[htbp]
  \centering
  \caption{Result of the subjective metric (\%)}
    \begin{tabular}{p{10.5em}cccccc}
    \toprule
    \textbf{Dataset\textbackslash{}Model} & \multicolumn{1}{p{5.6em}<{\centering}}{GeoGAN} & \multicolumn{1}{p{5.6em}<{\centering}}{Pix2Pix} & \multicolumn{1}{p{5.6em}<{\centering}}{CycleGAN} & \multicolumn{1}{p{5.6em}<{\centering}}{UNIT} & \multicolumn{1}{p{5.6em}<{\centering}}{Att-CycleGAN} & \multicolumn{1}{p{5.6em}<{\centering}}{SMAPGAN } \\
    \midrule
    A-10\% & 1.03 & 1.86 & 30.57 & 3.73 & 9.57 & \textit{\textbf{53.24}} \\
    A-50\% & 5.52 & 7.81 & 18.77 & 2.68 & 4.73 & \textit{\textbf{60.49}} \\
    B-10\% & 10.62 & 6.74 & 19.66 & 13.34 & 1.58 & \textit{\textbf{48.06}} \\
    B-50\% & 12.55 & 10.88 & 10.77 & 8.58 & 3.03 & \textit{\textbf{54.18}} \\
    \bottomrule
    \end{tabular}%
  \label{tab:addlabel}%
\end{table*}%

\subsection{Experiment Setup}
For all the experiments, we set $\lambda_{ctn}=10$, $\lambda_{adv}=1$, and $\lambda_{idt}=0.1$ in Formulation \eqref{eq1}, $\lambda_{L1u}=1$ in Formulation \eqref{eq3} and \eqref{eq4}, and $\lambda_{L1}=10$ in Formulation \eqref{eq5} and \eqref{eq6}. We adopted a 1-batch-size Adam optimizer \cite{pecka_data-driven_2018} with $\beta_1=0.5$, $\beta_2=0.999$ and a two-stage learning rate strategy used in \cite{zhu_unpaired_2017}. We trained the model for 200 epochs on each dataset and $t_s$ was set to 150. For metrics, in the SSIM (Formulation \eqref{eq26}) we set $C_1=C_2=1e^{-12}$. By contrast, in the ESSI (Formulation \eqref{eq20}) we set $C_1=C_2=1e^{-12}$. We both calculated the objective metrics by the luminance of the output and the mean value of that by three channels (R, G, B) to present the results globally. The training time of the model is presented in Table 1.

\subsection{Results}
\subsubsection{Qualitative Result}

Figure \ref{Figure 8} and Figure \ref{Figure 9} show qualitative comparisons among the five baselines and our SMAPGAN model. In comparison to GeoGAN and Pix2Pix, our model only trained by paired data performed better at generating structure features such as continuous color blocks and a regular border shape. In comparison to other four unsupervised models, our SMAPGAN, which was trained with all data did a better job of distinguishing different features and filling in the corresponding colors, in addition to obtaining a better structure similarity. Furthermore, the model trained on all data beat the one trained by only paired data and reached the best performance thanks to the semi-supervised learning strategy.In our experiment, we used same amounts of paired samples for SMAPGAN (only paired) and SMAPGAN. The difference between these two situations is that SMAPGAN has more unsampled training samples than SMAPGAN, which only has paired samples. In other words, SMAPGAN (only paired) applied a supervised strategy and SMAPGAN applied a semi-supervised strategy. The result shows the effectiveness of the semi-supervised strategy.

\begin{figure*}[htbp]
\centering
\includegraphics[scale=1]{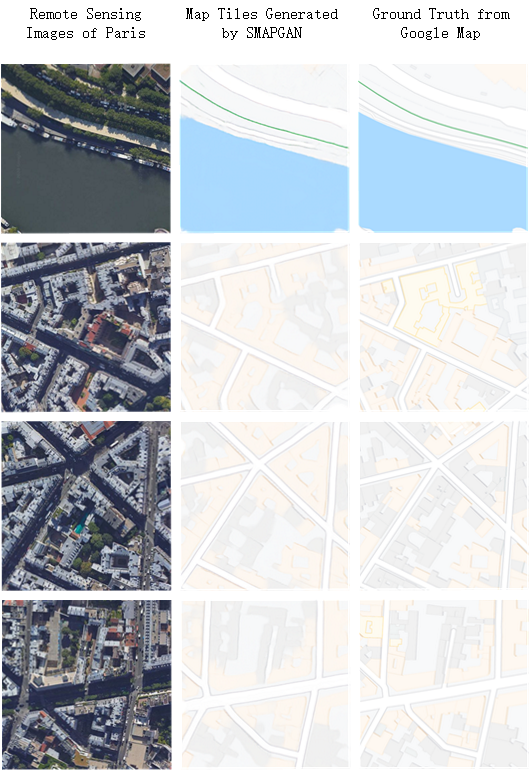}
\caption{Remote sensing images of Paris translated to map tiles }
\label{Figure 14}
\end{figure*}

In order to test whether the model can be applied to remote sensing images of other cities, we downloaded remote sensing images and map tiles of Paris from Google Maps at zoom 17 to operate the experiment. Figure \ref{Figure 14} shows the results, which demonstrate that remote sensing images of Paris can be translated to map tiles by our model, although some flaws exist because of shade on remote sensing images

\begin{figure*}[htbp]
\centering
\includegraphics[scale=0.45]{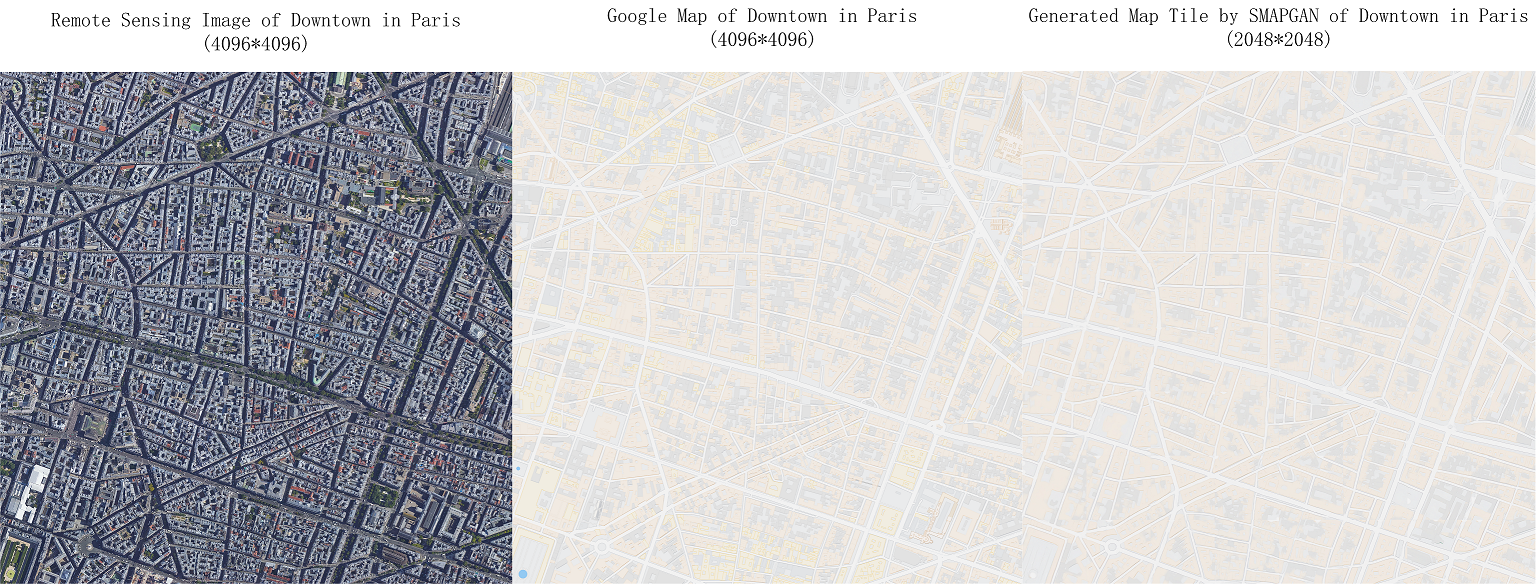}
\caption{A large scale remote sensing image translated to a map tile }
\label{Figure 15}
\end{figure*}

Also we attempted to translate a remote sensing image at a large scale, as shown in Figure \ref{Figure 15}. We downloaded a remote sensing image at a 4096*4096 scale and the corresponding map tile from Google Maps. Due to the input restriction of the model, we cropped the remote sensing image into 16 515*512 size images and resized each image to 256*256 to generate map tiles due to the limitation of the model for the size of the input. Then the model generated styled 256*256 map tiles and these map tiles were joined together to form a 2048*2048 map tile.


\subsubsection{Quantitative Result}
Similar to qualitative results, the results of the objective and subjective metrics show that our SMAPGAN outperforms the other baseline models. Table 2, 3, and 4 show an objective metrics comparison among all models, while Table 5 shows the results of the subjective metric.

First, we discuss the results of the objective metrics. For MSE, SMAPGAN performed worse slightly than GeoGAN. However, as discussed in 3.5, MSE can only measure the global mean distance of the pixel values between a generated map tile and its ground truth, which has a great flaw in this task.

Additionally, we can see that SMAPGAN outperforms the SOTA baselines in the SSIM and ESSI assessments. Specifically, a significant improvement of the ESSI is presented, which indicates an improvement in the topological relationship among the objects on generated map tiles by considering the topological consistency in our method.

Meanwhile, the model trained with a semi-supervised strategy obtained better results than that trained with only paired data and other unsupervised baselines, which proves the effectiveness of our semi-supervised strategy. In other words, the semi-supervised strategy takes the advantage of unpaired data to provide more valid information for learning, and takes the advantage of the paired data to offer an accurate target for training.

Then, we analyzed the results of the subjective metric. In the human perceptual test, every sample was assessed over 3 times on average and the number of participants was 145 person-time. For the 10\% and 50\% paired data on dataset A, our model gained 53.24\% and 60.49\% support, respectively. For the 10\% and 50\% paired data on dataset B, our model gained 48.06\% and 54.18\% support, respectively. Obviously, our model dominates the human perceptual test. This result reports that our SMAPGAN outperforms the SOTA baselines during human perceptual judgement.

\subsubsection{Threshold of Paired Samples Ratio} 
\begin{figure}[htbp]
\centering
\includegraphics[scale=0.3]{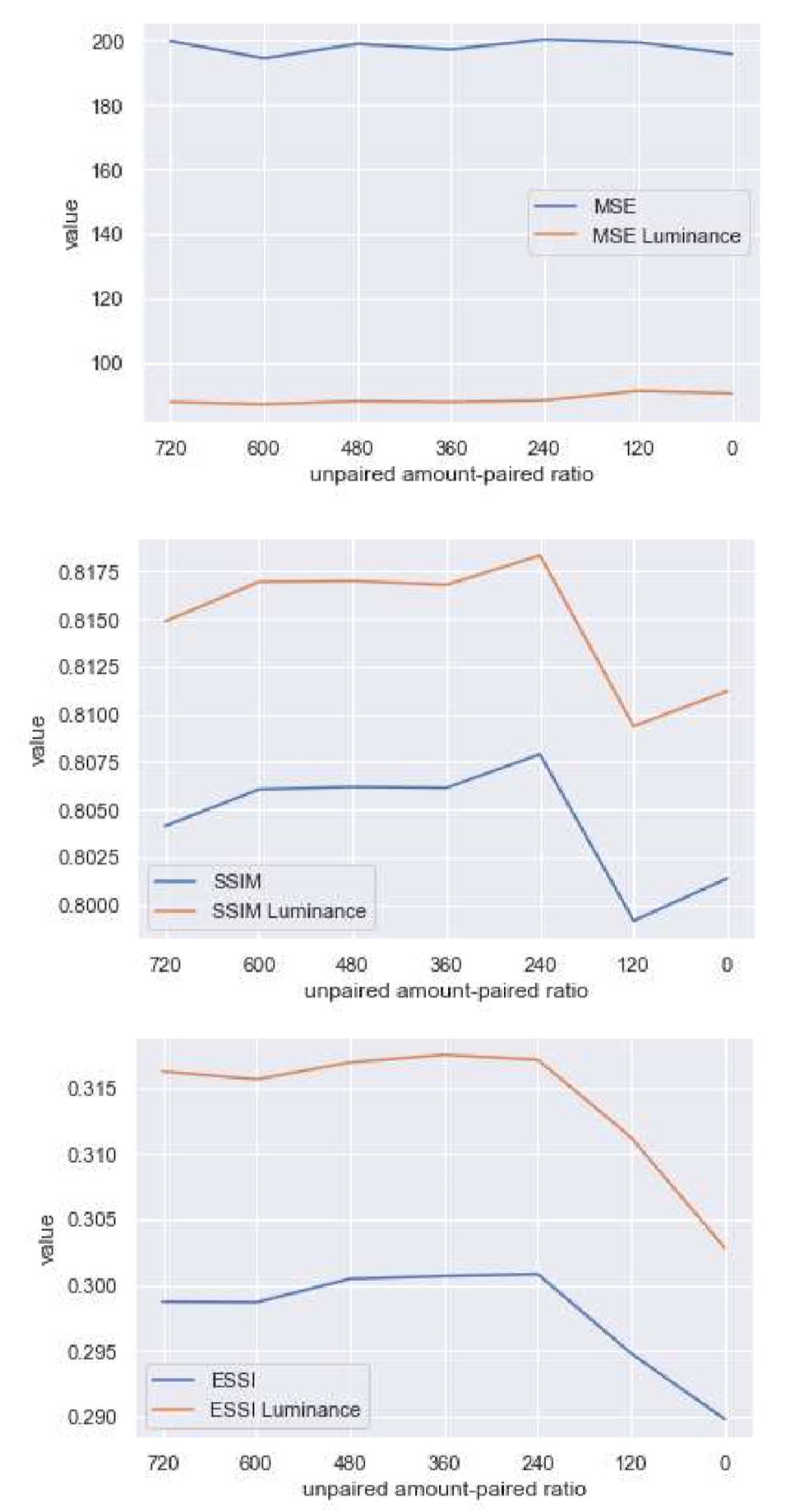}
\caption{The fluctuation in the measurement under decreasing paired samples  }
\label{Figure 12}
\end{figure}

\begin{figure}[htbp]
\centering
\includegraphics[scale=0.3]{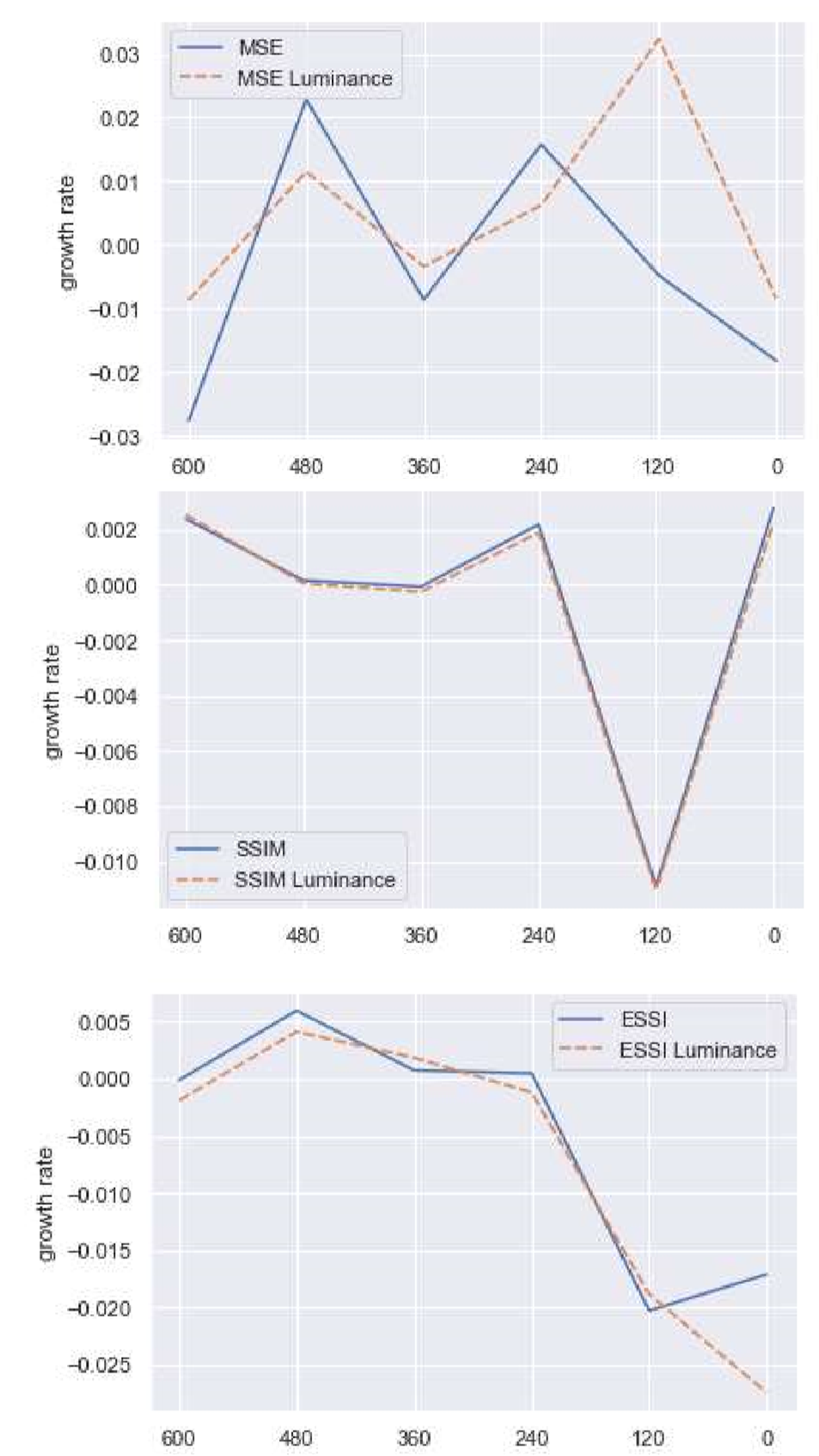}
\caption{The rate of increase in the measurement under decreasing paired samples}
\label{Figure 13}
\end{figure}

In order to find the threshold of the amount of paired samples, we performed an experiment, with the results shown in Figure \ref{Figure 12} and Figure \ref{Figure 13}. The experiment fixed the number of paired samples at 360 pairs (720 images), and decreased the amount of unpaired samples from 1440 (720 remote sensing images and 720 map tiles) to increase the ratio of paired samples in the training set. The ratio of the paired samples was increased from 1/3 (720 paired images and 1440 unpaired images) to 1 (720 paired images and 0 unpaired images). These samples were selected randomly from dataset A. We repeated this experiment five times and obtain an average result, as shown in Figure \ref{Figure 12} and Figure \ref{Figure 13}.

Figure \ref{Figure 12} shows that increasing the paired ratio does not affect the MSE very significantly. However, for the SSIM and ESSI, peak points obviously appear. When unpaired samples are 240 pair (240 remote sensing images and 240 map tiles, paired samples ratio is 60\%), the measurement began to decrease. Figure \ref{Figure 13} also proves that more accurately. When the paired ratio is 60\%, the increase rate of the measurements began to drop negative and when the paired ratio is larger than 1/3 (where the unpaired samples are more than 240 pairs), the increase rate of the ESSI and SSIM can be positive while the increase rate of the MSE fluctuates slightly. We thought that the best paired ratio for the performance of the model is between 1/3 and 60\%. When paired samples ratio is more than 60\%, adding up unpaired samples in training leads to an unstable increase rate of the measurement. When the paired ratio is less than 1/3, the MSE and ESSI are negative. Restricted by the number of samples, this result might not be very accurate.

\subsection{Ablation Study}
The content loss consists of three parts: L1 loss, image gradient L1 loss and image gradient structure loss. The experimental results show that our model with a content loss, including L1 loss, image gradient L1 loss and image gradient structure loss, performed better than models using a content loss with only L1 loss (CycleGAN and Pix2pix). In order to discuss the impact of different losses on the generated map tiles, we conducted an ablation study.

\begin{figure*}[htbp]
\centering
\includegraphics[scale=0.65]{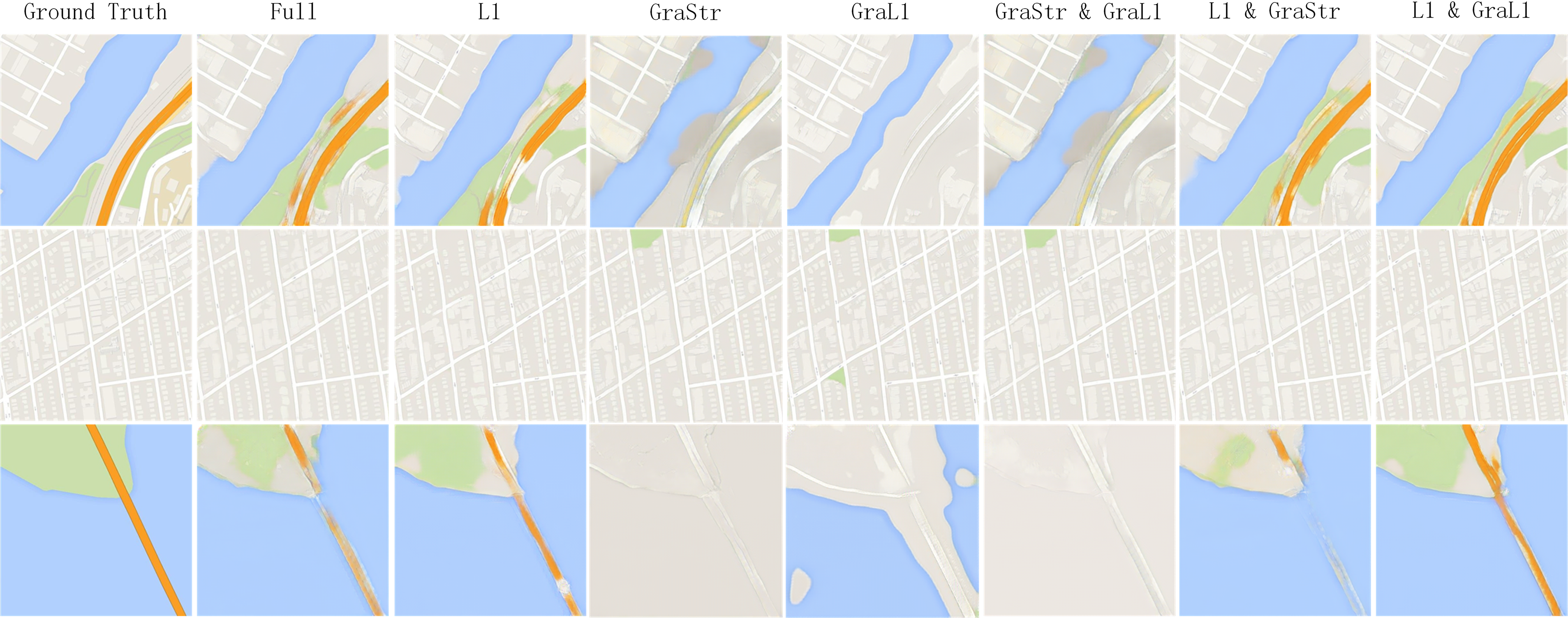}
\caption{Three selected samples from the results of the ablation study conducted on dataset A-50\%. Each sample included the ground truth and the map tiles generated by models trained with a full loss function, the L1 loss alone, the proposed structure constraint alone, the L1 loss and Gradient structure loss, and the L1 loss and Gradient L1 loss.}
\label{Figure 10}
\end{figure*}

We compared against the ablations of the content loss on dataset A-50\%. Figure 10 shows the qualitative effect of the three components of the content loss. The structure quality, such as color-continuity is not good when L1 Loss is used alone, which is adopted in universal img2img work. We then tried only using the GraStr Loss and GraL1 Loss proposed in this work when training our model. In this case, the model only focuses on generating map tiles that have similar objects outlined as ground truths, but it does not penalize the mismatch on feature colors between them, which is the symbol of types of object features. Therefore, erroneous object classification happens, although the topological relationships of objects on generated map tiles are more similar to the real scenario. To further determine how the two specific types of topological consistency losses work, we conducted a test with each of the two losses removed. We found that removing the GraL1 Loss results in a bad color continuity and blurring outline, whereas disabling the GraStr Loss causes the model to fail at generating maps with correct topological relationships of the objects. The phenomenon claims that the image's GraL1 Loss and GraStr Loss guide the model to learn pixel-level continuity and topological consistency, respectively.

We performed a quantitative ablation study for various loss combinations. The results are presented in Tables 6, 7, and 8 and we only present the results of luminance to the simplify analysis.

\begin{table}[h]
  \centering
  \caption{Comparison of objective metrics of L1 loss's ablation study.}
    \begin{tabular}{p{7.5em}ccc}
    \toprule
    \multicolumn{1}{p{7.5em}<{\centering}}{\textbf{Component\textbackslash{}Metric}} & \multicolumn{1}{p{4.19em}<{\centering}}{\textbf{MSE}} & \multicolumn{1}{p{4.19em}<{\centering}}{\textbf{SSIM}} & \multicolumn{1}{p{4.19em}<{\centering}}{\textbf{ESSI}} \\
    \midrule
    Full & 83.77 & 0.8105 & 0.3234 \\
    L1   & 95.80 & 0.7915 & 0.2956 \\
    \textit{L1\&GraStr} & \textit{\textbf{83.97}} & \textit{\textbf{0.8105}} & \textit{\textbf{0.3227}} \\
    L1\&GraL1 & 90.11 & 0.7942 & 0.3034 \\
    \bottomrule
    \end{tabular}%
  \label{tab:addlabel}%
\end{table}%

Table 6 shows the results of the ablation study of the L1 loss, which indicates that the GraStr Loss and GraL1 Loss contribute to increasing all the objective metrics. Furthermore, GraStr outperforms GraL1 to improve the global quality of the generated map tiles. GraStr improves the SSIM and ESSI scores significantly, which implies that GraStr greatly helps the generator improve the topological relationship of objects on generated map tiles. This improvement enhances global pixel-wise consistency, which the drop in the MSE score represents. Meanwhile, the slight improvement of all metrics after adding GraL1 shows that GraL1 fine tuned the quality of the generated map tiles.

\begin{table}[h]
  \centering
  \caption{Comparison of the objective metrics of image gradient L1 loss's ablation study.}
    \begin{tabular}{p{7.5em}ccc}
    \toprule
    \multicolumn{1}{p{7.5em}<{\centering}}{\textbf{Component\textbackslash{}Metric}} & \multicolumn{1}{p{4.19em}<{\centering}}{\textbf{MSE}} & \multicolumn{1}{p{4.19em}<{\centering}}{\textbf{SSIM}} & \multicolumn{1}{p{4.19em}<{\centering}}{\textbf{ESSI}} \\
    \midrule
    Full & 83.77 & 0.8105 & 0.3234 \\
    GraL1 & 210.39 & 0.7300 & 0.2379 \\
    GraStr\& GraL1 & 193.66 & 0.7999 & 0.3229 \\
    \textit{L1\&GraL1} & \textit{\textbf{90.11}} & \textit{\textbf{0.7942}} & \textit{\textbf{0.3034}} \\
    \bottomrule
    \end{tabular}%
  \label{tab:addlabel}%
\end{table}%

Table 7 shows the results of the ablation study of the image gradient L1 loss. Compared to Table 5, the single GraL1 loss performs very poorly under all objective metrics. When GraStr was combined with GraL1, the SSIM and ESSI increased greatly and MSE increased slightly, which proves the contribution of GraStr in improving the topological relationship of objects on generated map tiles, as has been discussed for Table 5. The L1 loss improves the global pixel-wise consistency, which the drop in the MSE shows. This improvement contributes to the rise of the SSIM and ESSI. Thus, L1 is a vital loss for guiding the generator to maintain pixel-wise consistency, which is a primary constraint for the map translation task.

\begin{table}[h]
  \centering
  \caption{Comparison of the objective metrics of the image gradient structure loss's ablation study.}
    \begin{tabular}{p{7.5em}ccc}
    \toprule
    \multicolumn{1}{p{7.5em}<{\centering}}{\textbf{Component\textbackslash{}Metric}} & \multicolumn{1}{p{4.19em}<{\centering}}{\textbf{ MSE }} & \multicolumn{1}{p{4.19em}<{\centering}}{\textbf{ SSIM }} & \multicolumn{1}{p{4.19em}<{\centering}}{\textbf{ ESSI }} \\
    \midrule
    Full & 83.77 & 0.8105 & 0.3234 \\
    GraStr & 162.81 & 0.7996 & 0.3222 \\
    GraStr\&GraL1 & 193.66 & 0.7999 & 0.3229\\
    \textit{L1\&GraStr} &\textit{\textbf{ 83.97}} & \textit{\textbf{ 0.8105}} & \textit{\textbf{ 0.3227}} \\
    \bottomrule
    \end{tabular}%
  \label{tab:addlabel}%
\end{table}%
\begin{figure*}[htbp]
\centering
\includegraphics[scale=0.7]{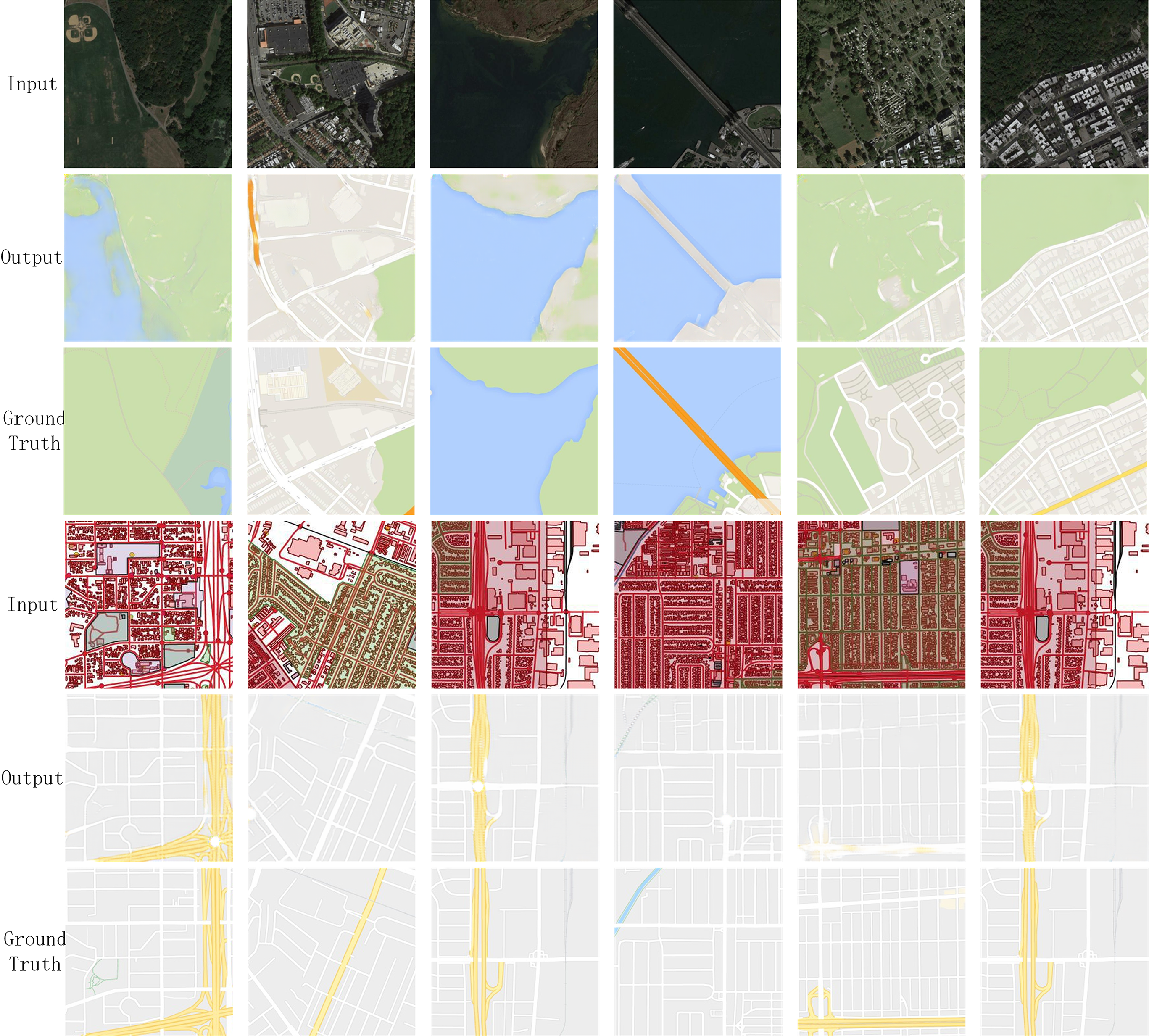}
\caption{ Typical remaining errors.}
\label{Figure 11}
\end{figure*}

Table 8 shows the results of the ablation study of the image gradient structure loss. A single GraStr obtains a high SSIM and ESSI, which indicates the importance of GraStr to improve the topological relationship of objects on generated maps. However, a high MSE shows the flaw of the GraStr to maintain a pixel-wise consistency, which can be improved by adding an L1 loss. Meanwhile, by adding GraL1, all the objective metrics improve slightly, again proving its fine tuning function.

In conclusion, the ablation study shows different effects of the L1 loss, GraStr loss, and GraL1 loss. The L1 loss and GraStr loss are important for the generator because the L1 loss constrains the generator to decrease the global pixel-wise distance between the output and the ground truth and the GraStr loss constrains the generator to improve the topological consistency between the output and the ground truth. The GraL1 loss can fine tune the quality of generated map tiles under the constraint of an L1 loss and GraStr loss. The quantitative results are consistent with the qualitative results.

\subsection{Discussion}

SMAPGAN provides a rapid method of generating map tiles directly from remotely sensed images with pixel-to-pixel translating. It is surprising that data from completely different domains, such as remotely sensed images and map tiles, can be converted very well under a deep generative network framework. There are still many imperfections though, which has inspired us to explore more possibilities as follows:

(1) Combining the edge information of objects in remotely sensed images can possibly improve the SMAPGAN's performance. We have observed some breaks in the roads in Figure \ref{Figure 8}. We speculate that the unbroken edge information can benefit from keeping the globality of objects, such as roads.

(2) Combining additional vector information may also help improve SMAPGAN's performance and generate map tiles with semantic control.

(3) Transforming remotely sensed images and map tiles with different resolutions and multiple styles.

SMAPGAN also shows its potential to generate map tiles in multiple ways; not just remotely sensed images but also labels, or even styled painting. It is a renewed way to product map tiles rapidly for many fields such as art maps, games, and even planning and design.
\section{Conclusion}
In this work, we put forward a novel semi-supervised structure augmented model, called SMAPGAN, to automatically generate styled map tiles. This strategy, which is pre-trained with unpaired data and fine-tuned with paired data, shows the potential to generate more real styled map tiles with a small number of paired samples. We also show that topological consistency is very important for improving the fidelity of generated styled map tiles. The designed image gradient L1 loss (GraL1) and the image gradient structure loss (GraStr) are simple but effective loss functions for capturing the topological consistency of generated styled maps. Moreover, a new full-reference metric, ESSI, which focuses on the quality of the topological relationship of objects on generated map tiles is proposed to assess topological consistency between generated map tiles and their ground truths. The experiments show that SMAPGAN outperforms the SOTA baselines in terms of both objective metrics and subjective human perceptual evaluation, which shows the effectiveness of our model. In future works, we will focus on erroneous objects translation, wrong topological relations and shade problems in generated maps.
\appendix[Proof of ESSI]
\subsection{Topological Consistency Measurement Through the Pearson Coefficient} 
We hope to find a method to measure whether a generated map tile can replicate the topological relationship among objects on the ground truth. As discussed in section 3.3, a topological relationship among the pixels of the edge of objects can reflect a topological relationship among objects. Thus, we apply the Canny edge detector \cite{canny_computational_1986} to extract edge images of the generated map and its ground truth, as shown in Figure \ref{Figure 3}.

Thus, we have two pixel matrices to measure their distance. We need a measurement to measure the distance considering the coordinate of each pixel because each point in the pixel value matrix has coordinates for the row and column numbers, which preserves its topological relationship with other points.

We dimensionally reduced their M$\times$ N pixel value matrices into M$\times$ N-dimension vectors via the progressive scanning method. We denote the reduced vector of the edge image of a generated map tile as $\mathcal{E}\left(G\left(x_R\right)\right)$ and the reduced vector of the edge image of the ground truth as $\mathcal{E}\left(x_M\right)$.

After dimensional reduction, the reduced vector maintains the topological relationship among points because each coordinate of points would only be mapped to one dimension of the vector, and each dimension of the vector is mapped to only one coordinate. Thus, the topological consistency between the generated map tiles and the ground truth is transformed to the correlation between $\mathcal{E}\left(G\left(x_R\right)\right)$ and $\mathcal{E}\left(x_M\right)$. The absolute value of the Pearson correlation coefficient between these two edge images is employed to measure the correlation:

\begin{equation}
    \begin{aligned}
    \label{eq21}
    \rho(&G(x_R),x_M)=\frac{|\sigma_{\mathcal{E}(G(x_R))\mathcal{E}(x_M)}|+C_1}{\sigma_{\mathcal{E}(G(x_R))}\sigma_{\mathcal{E}(x_M)}+C_1}
    \end{aligned}
\end{equation}

Where $\sigma_{\mathcal{E}\left(G\left(x_R\right)\right)}$ and $\sigma_{\mathcal{E}\left(x_M\right)}$ are the standard deviation of $\mathcal{E}\left(G\left(x_R\right)\right)$ and $\mathcal{E}\left(x_M\right)$, respectively; $\sigma_{\mathcal{E}\left(G\left(x_R\right)\right)\mathcal{E}\left(x_M\right)}$ is the covariance between $\mathcal{E}\left(G\left(x_R\right)\right)$ and $\mathcal{E}\left(x_M\right)$; and $C_1$ is a constant to keep $\rho\left(G\left(x_R\right),x_m\right)$ stable when $\sigma_{\mathcal{E}\left(G\left(x_R\right)\right)}\sigma_{\mathcal{E}\left(x_M\right)}$ is close to 0.

\subsection{Avoid a Wrong Evaluation by Adding GMS} 
\begin{figure}[ht]
\centering
\includegraphics[scale=0.6]{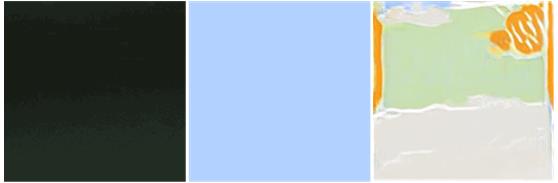}
\caption{From the left to right are the remote sensing image of the sea, its corresponding Google Map tile, and the incorrectly generated map. Obviously, the generated map tile is poor but $\rho\left(G\left(x_R\right),x_m\right)$ is very high because the edge matrix of the map tile is an 0 matrix.}
\label{Figure 4}
\end{figure}

However, in our experiment, an overestimation would occur if we only use $\rho\left(G\left(x_R\right),x_m\right)$ to evaluate the generated map tile. As shown in Figure \ref{Figure 4}, if $G\left(x_R\right)$ is wrongly generated such that $\mathcal{E}\left(G\left(x_R\right)\right)\neq$0 while $\mathcal{E}\left(x_M\right)=0$, then $\sigma_{\mathcal{E}\left(G\left(x_R\right)\right)\mathcal{E}\left(x_M\right)}$ and $\sigma_{\mathcal{E}\left(G\left(x_R\right)\right)}\sigma_{\mathcal{E}\left(x_M\right)}$ equal to 0, but $\rho\left(G\left(x_R\right),x_m\right)$ equals to 1. Thus, $G\left(x_R\right)$ would be wrongly evaluated to be very good. To avoid this case, we constructed a coefficient that is inspired by the gradient magnitude similarity (GMS) proposed in \cite{xue_gradient_2014}:

\begin{equation}
    \begin{aligned}
    \label{eq22}
    \theta(&G(x_R),x_M)=&\frac{2\mu_{\mathcal{E}(G(x_R))}\mu_{\mathcal{E}(x_M)}+C_2}{\mu^2_{\mathcal{E}(G(x_R))}+\mu^2_{\mathcal{E}(x_M)}+C_2}
    \end{aligned}
\end{equation}

Where $\mu_{\mathcal{E}\left(G\left(x_R\right)\right)}$ and $\mu_{\mathcal{E}\left(x_M\right)}$ are the mean values of $\mathcal{E}\left(G\left(x_R\right)\right)$ and $\mathcal{E}\left(x_M\right)$, respectively. Thus, when $\mathcal{E}\left(x_M\right)$ is an 0 matrix and $\mathcal{E}\left(G\left(x_R\right)\right)$ is not, we have the following:

\begin{equation}
    \begin{aligned}
    \label{eq23}
    \sigma_{\mathcal{E}(G(x_R))\mathcal{E}(x_M)}=\sigma_{\mathcal{E}(G(x_R))}\sigma_{\mathcal{E}(x_M)}&=0\\
    \rho(G(x_R),x_M)&=1\\
    2\mu_{\mathcal{E}(G(x_R))}\mu_{\mathcal{E}(x_M)}&=0\\
    \mu^2_{\mathcal{E}(G(x_R))}+\mu^2_{\mathcal{E}(x_M)}&=\mu^2_{\mathcal{E}(G(x_R))}\\
    \theta(G(x_R),x_M)=\frac{2\mu_{\mathcal{E}(G(x_R))}\mu_{\mathcal{E}(x_M)}+C_2}{\mu^2_{\mathcal{E}(G(x_R))}+\mu^2_{\mathcal{E}(x_M)}+C_2}&=\frac{C_2}{\mu^2_{\mathcal{E}(G(x_R))}+C_2}
    \end{aligned}
\end{equation}

By setting $C_2$ appropriately, we can make $\theta$ very close to 0. When $\mathcal{E}\left(x_M\right)=0$, and $G\left(x_R\right)$ is well generated, then $\mathcal{E}\left(G\left(x_R\right)\right)=0$, and $\theta=1$. Then, we can obtain the ESSI:

\begin{equation}
    \begin{aligned}
    \label{24}
    ESSI(G(x_R),x_M)= \theta(G(x_R),x_M)\rho(G(x_R),x_M)\\=\frac{(|\sigma_{\mathcal{E}(G(x_R))\mathcal{E}(x_M)}|+C_1)(2\mu_{\mathcal{E}(G(x_R))}\mu_{\mathcal{E}(x_M)}+C_2)}{(\sigma_{\mathcal{E}(G(x_R))}\sigma_{\mathcal{E}(x_M)}+C_1)(\mu^2_{\mathcal{E}(G(x_R))}+\mu^2_{\mathcal{E}(x_M)}+C_2)}
    \end{aligned}
\end{equation}

This metric is symmetric, such that $ESSI\left(G\left(x_R\right),x_M\right)=ESSI\left(x_M,G\left(x_R\right)\right)$, bounded where $ESSI\in\left[0,1\right]$, and $ESSI=1$ if and only if the $M\times N$ dimension vector $\mathcal{E}\left(G\left(x_R\right)\right)=\ \mathcal{E}\left(x_M\right)$, and $C\neq0$.
\section*{Acknowledgment}
The authors would like to thank the funding and support from the Advance Research Projects of Civil Aerospace Technology, Intelligent Distribution Technology of Domestic Satellite Information, No. B0301 and the National Science Foundation of China (grant numbers 41871364, 41571397, 41871276, and 51678077). Thanks to T. Xu's granny Ms. Mengjun Yan, R.I.P.

\ifCLASSOPTIONcaptionsoff
  \newpage
\fi



\bibliographystyle{IEEEtran}
\bibliography{smapganreference.bib}
%



%
\begin{IEEEbiography}[{\includegraphics[width=1in,height=1.25in,clip,keepaspectratio]{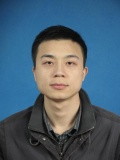}}]{Xu~Chen}
is currently an associate professor in the School of Computer Science at Wuhan University. He was born in 1982. He received his Bachelor degree from the School of Computer Science at Wuhan University in 2004 and his Ph.D. degree from the State Key Laboratory of Information Engineering in Surveying, Mapping and Remote Sensing at Wuhan University in 2010. His main research interests cover Spatial Information Processing and Information Retrieval.
\end{IEEEbiography}
\begin{IEEEbiography}[{\includegraphics[width=1in,height=1.25in,clip,keepaspectratio]{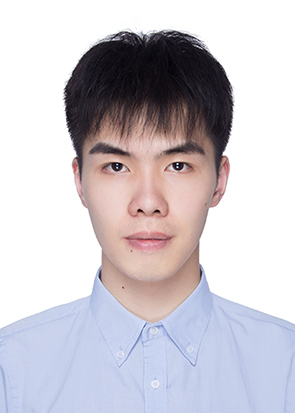}}]{Songqiang~Chen}
is currently an undergraduate student in the school of computer science, Wuhan University, Wuhan, China, where he would be a postgraduate student soon. He focuses on deep learning, software engineering, software testing, natural language processing and computer vision.
\end{IEEEbiography}
\begin{IEEEbiography}[{\includegraphics[width=1in,height=1.25in,clip,keepaspectratio]{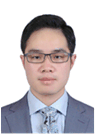}}]{Tian~Xu}
recieved the bachelor's degree in mathematics base class from the Wuhan University, Wuhan, China, in 2013, and the master's degree in quantitative finance from the University of Glasgow, Glasgow, UK, in 2017. He is currently a Research Associate in the school of computer science, Wuhan University, Wuhan, China, who works on machine learning and financial statistics.
\end{IEEEbiography}
\begin{IEEEbiography}[{\includegraphics[width=1in,height=1.25in,clip,keepaspectratio]{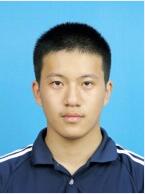}}]{Bangguo~Yin}
is an undergraduate majoring Software Engineering in the School of Computer Science, Wuhan University, Wuhan, China.
\end{IEEEbiography}
\begin{IEEEbiography}[{\includegraphics[width=1in,height=1.25in,clip,keepaspectratio]{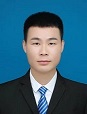}}]{Jian~Peng}
is a PhD candidate in the cartography and geographical information system from Central South University Changsha, China. His research topic is the image intelligent understanding inspired by brain memory mechanism. His research interests include neural computation, continual learning and remote sensing.
\end{IEEEbiography}
\begin{IEEEbiography}[{\includegraphics[width=1in,height=1.25in,clip,keepaspectratio]{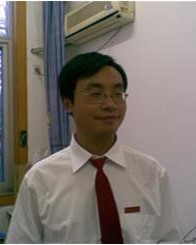}}]{Xiaoming~Mei}
received the B.S. degree in Applied Geophysics from Central South University, Changsha, China, in 2000.Received the master’s degree in Geo-detection and Information Technology from Central South University, Changsha, China, in 2003. And received the Ph.D.degree in Photogrammetry and Remote Sensing from Wuhan University, China, in 2007.He is currently an Instructor with the School of Geosciences and Info-Physics, Central South University, Changsha, China. His current research interests include multi- and hyper-spectral remote sensing data processing, high resolution image processing and scene analysis, LiDAR data processing, and computational intelligence.
\end{IEEEbiography}
\begin{IEEEbiography}[{\includegraphics[width=1in,height=1.25in,clip,keepaspectratio]{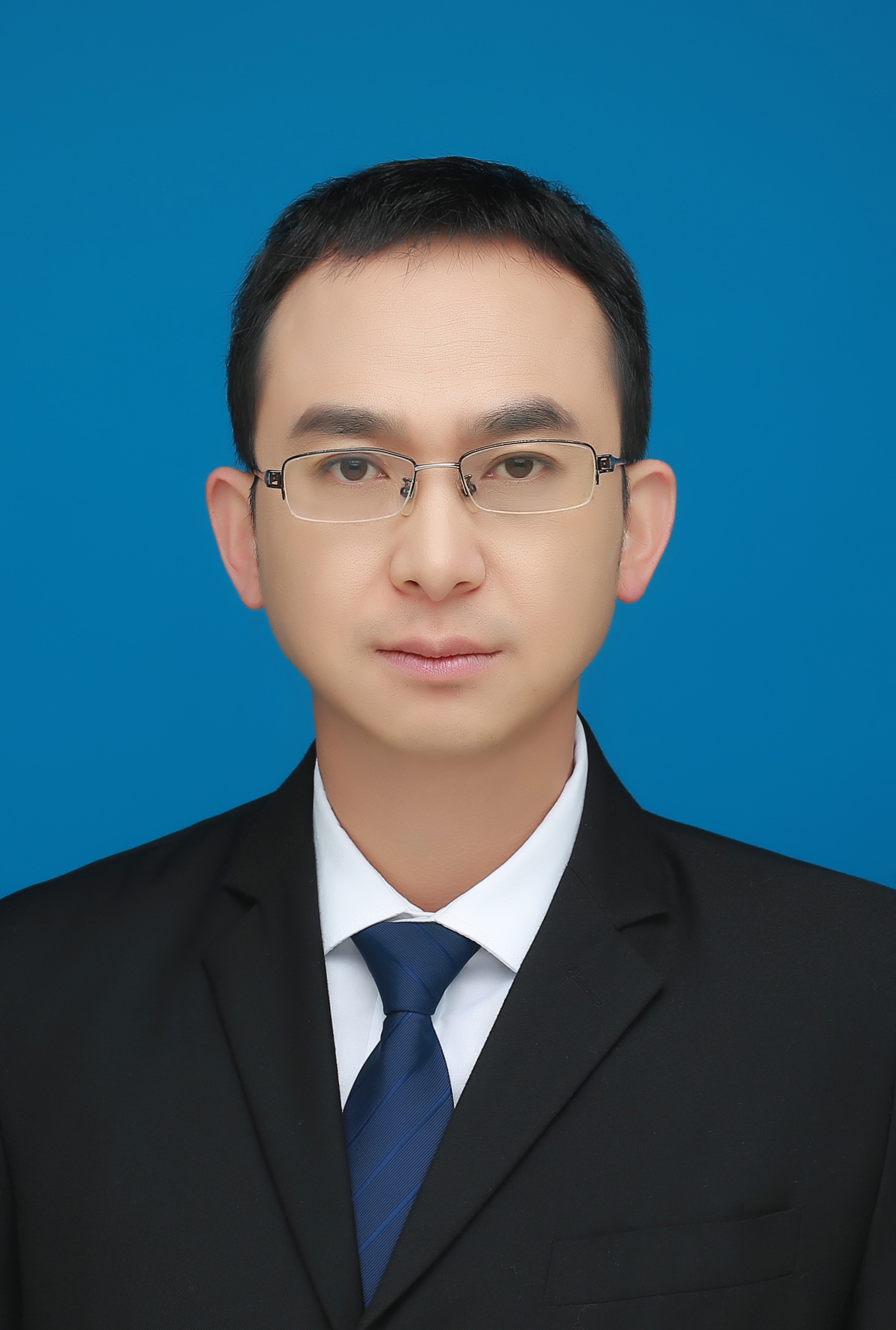}}]{Haifeng~Li}
received the master's degree in transportation engineering from the South China University of Technology, Guangzhou, China, in 2005, and the Ph.D. degree in photogrammetry and remote sensing from Wuhan University, Wuhan, China, in 2009. He is currently a Professor with the School of Geosciences and Info-Physics, Central South University, Changsha, China. He was a Research Associate with the Department of Land Surveying and Geo-Informatics, The Hong Kong Polytechnic University, Hong Kong, in 2011, and a Visiting Scholar with the University of Illinois at Urbana-Champaign, Urbana, IL, USA, from 2013 to 2014. He has authored over 30 journal papers. His current research interests include geo/remote sensing big data, machine/deep learning, and artificial/brain-inspired intelligence. He is a reviewer for many journals.
\end{IEEEbiography}






\end{document}